\documentclass[12pt,preprint]{aastex}

\usepackage{epsfig}

\slugcomment{}

\shorttitle{STIS spectroscopy of ULIRGs}
\shortauthors{Farrah et al.}

\begin{document}

\title{STIS ultraviolet/optical spectroscopy of `warm' ultraluminous infrared galaxies}

\author{D. Farrah \& J. A. Surace}
\affil{Spitzer Science Center, Jet Propulsion Laboratory, California Institute of Technology, Pasadena, CA 91125}
\author{S. Veilleux}
\affil{University of Maryland, Department of Astronomy, College Park, MD 20742}
\author{D. B. Sanders}
\affil{Institute for Astronomy, University of Hawaii, Honolulu, Hawaii; and Max-Planck-Institut fur extraterrestrische Physik, D-85740 Garching, Germany}
\author{W. D. Vacca}
\affil{SOFIA-USRA, NASA Ames Research Center, MS144-2, Moffett Field, CA 94035}

\begin{abstract}
We present high spatial resolution ultraviolet and optical spectroscopy, obtained using the Space Telescope Imaging 
Spectrograph on board the Hubble Space Telescope, of nuclear structures within four `warm' Ultraluminous Infrared Galaxies 
(ULIRGs). We find an AGN in at least three, and probably all four of our sample, hosted in a compact, optically luminous 
`knot'. In three cases these knots were previously identified as a putative AGN nucleus from multiband optical imaging. 
Three of the sample also harbor a starburst in one or more knots, suggesting that the optically luminous knots seen in 
local ULIRGs are the most likely sites of the dust-shrouded starburst and AGN activity that power the infrared emission. 
The four AGN have a diverse range of properties; two are classical narrow line AGN, one shows both broad and narrow lines 
and evidence for lines of sight from the narrow through to the broad line regions, and one is plausibly a FeLoBAL AGN. The 
probable presence in one object of an FeLoBAL AGN, which are extremely rare in the QSO population, supports the idea that 
LoBAL AGN may be youthful systems shrouded in gas and dust rather than AGN viewed along a certain line of sight. The three 
starbursts for which detailed constraints are possible show a smaller range in properties; all three bursts are young with 
two having ages of $\sim4$Myr and the third having an age of 20Myr, suggesting that ULIRGs undergo several bursts of star 
formation during their lifetimes. None of the starbursts show evidence for Initial Mass Function slopes steeper than about 
3.3. The metallicities of the knots for which metallicities can be derived are all at least 1.5$Z_{\odot}$. The properties 
of one further starburst knot are consistent with it being the forming core of an elliptical galaxy. Our results suggest 
that detailed studies of the `knots' seen in ULIRGs can give important insights into the most violent starburst and AGN 
activity at both low and high redshift.
\end{abstract}


\section{Introduction}
Ultraluminous Infrared Galaxies (ULIRGs, those objects with 1-1000$\mu$m luminosities in excess of $10^{12}L_{\odot}$) were first 
discovered by \citet{rie}, and have been intensely studied since the discovery by the Infrared Astronomical Satellite (IRAS) of 
significant populations of ULIRGs in the local Universe \citep{soi1}. Unveiling the power source for the infrared (IR) emission 
in ULIRGs, thought to be some combination of starburst and AGN activity triggered by ongoing interactions (see \citet{san} for a 
review), is important for two reasons. Firstly, local ULIRGs, although rare, represent the closest examples of the most extreme 
cases of star formation and black hole accretion in galaxies, and can be used to study in detail how such processes drive galaxy 
evolution. Secondly, the ULIRG luminosity function shows strong evolution with redshift \citep{sau} and deep sub-mm 
surveys \citep{hug,eal,scot,bor} have found that systems with ULIRG-like luminosities are much more numerous at $z\gtrsim1$ than 
locally. A detailed understanding of local ULIRGs can therefore give insights into the power source \citep{rr2000,ver,far4,far5} 
and the role of interactions \citep{far3,cha03,consa,cons} in distant IR-luminous sources, and the impact of these sources on the 
cosmic history of star formation.

After just over three decades of study, a consensus has begun to emerge on the origin of the IR emission in ULIRGs. Early studies 
using optical spectroscopy showed that some ULIRG spectra resembled those of starburst galaxies \citep{jow}, whereas others contained emission 
lines characteristic of Seyferts \citep{san2}. Radio observations showed direct evidence for dust shrouded starbursts in some ULIRGs 
\citep{con1,smi1} and AGN cores in others \citep{lon2003,nag}. Papers based on mid-infrared spectroscopy suggested that most ULIRGs 
($\sim80\%$) are powered by starbursts \citep{rig,gen}, but that at least half of local ULIRGs contain both a starburst and an AGN. 
Results from modelling the 1-1000$\mu$m spectral energy distributions of local ULIRGs \citep{far2} show that all ULIRGs contain a 
dusty starburst, but that only about half contain an AGN that contributes significantly to the total IR emission. These results 
are consistent with X-ray observations of small samples of ULIRGs, which indicate that the bolometric emission from most ULIRGs is 
dominated by starburst activity rather than an AGN \citep{fra03,pta03}. There is also evidence that the presence of an AGN in ULIRGs 
depends on the total IR luminosity; optical and near-IR spectroscopic surveys \citep{vei,vei2} show a sharp rise in the fraction of 
ULIRGs with AGN signatures with increasing IR luminosity, and mid-IR spectroscopy from ISO \citep{tra} showed that ULIRGs with 
$L_{ir}<10^{12.4}L_{\odot}$ were mostly starburst dominated, whereas ULIRGs with $L_{ir}>10^{12.4}L_{\odot}$ were more likely to 
contain an AGN. 

Although we now have a much clearer picture of the role of starbursts and AGN in ULIRGs, few studies have, to date, correlated 
starburst and AGN activity in ULIRGs with morphological features (as opposed to correlating the power source in ULIRGs with broad scale 
morphologies, see e.g. \citet{vei3}), due mainly to the poor spatial resolution of current mid- and far-IR instruments when compared 
to the sub-arcsecond scale structures seen in UV/optical images of local ULIRGs. Correlating power source with morphological features 
in ULIRGs however has the potential to be a powerful probe of their nature and evolution;  most ULIRGs display a plethora of tails and 
loops, and compact `knots' in the nuclear regions \citep{sur,far1,bus2002}, and the `knots' in particular are suggested by some 
observations as possible sites for the dust-shrouded starburst and AGN activity in ULIRGs \citep{sur,soi2}.

A subset of the ULIRG population that is well suited to exploring links between morphology and power source are those ULIRGs with 
`warm' infrared colors ($f_{25}/f_{60}>0.2$)\footnote{The quantities $f_{12}$, $f_{25}$, $f_{60}$ \& $f_{100}$ represent the non color 
corrected IRAS flux densities in units of Janskys at 12$\mu$m 25$\mu$m 60$\mu$m \& 100$\mu$m respectively}. This color selection 
favors ULIRGs that contain dust at temperatures $\gtrsim50$K, which is most plausibly attributed to an AGN rather than a starburst. 
Coupled with the apparent ubiquity of starbursts in ULIRGs \citep{far2}, this color selection is likely to include those ULIRGs that 
contain both a starburst and an AGN, where the interplay between these two processes can be studied in detail. In this paper, we use 
the Space Telescope Imaging Spectrograph (STIS) on board the Hubble Space Telescope (HST) to obtain long-slit ultraviolet 
and optical spectra of several `knots' within the nuclear regions of four `warm' ULIRGs. Observations and analysis are described in 
\S\ref{obsanl} and results are given in \S\ref{results}. Our conclusions are presented in \S\ref{conc}. We assume $\Omega=1$, 
$\Lambda=0.7$ and $H_{0}=65$ km s$^{-1}$ Mpc$^{-1}$. Luminosities are quoted in units of bolometric solar luminosities, where 
$L_{\odot} = 3.826\times10^{26}$ Watts.

\section{Observations \& analysis} \label{obsanl}
We selected for observation ULIRGS from the sample with `warm' mid-infrared colors imaged in the optical using the Wide Field Planetary 
Camera 2 (WFPC2) on board HST by \citet{sur} and with adaptive optics by \citet{sur2}. This imaging demonstrated the presence of 
compact ($<$100 pc) high surface brightness features in the galaxy cores. Furthermore, these `warm' systems show evidence for Seyfert 
activity, which led \citet{san2} to postulate that these were objects transitioning between the general ULIRG population and optically 
selected quasars. Increasing central concentration in the infrared, as well as optical colors and luminosities, led \citet{sur} and 
\citet{sur2} to further identify some of these knots as possible active nuclei.

The four objects that were observed (Table \ref{sample}) span a range in properties, including both red (F12071-0444) and blue 
(F01003-2238) colors, simple (F01003-2238) and complex (F15206+3342) nuclear morphologies, single (F01003-2238, F12071-0444) and 
double (F05189-2524) nucleus systems, and HII, Sy 2, and Sy 1.5 spectral types. The data were taken in cycle 8 (proposal ID 8190) 
with STIS in longslit spectroscopy mode on board the HST. Three spectra were taken of each object, with the $52\arcsec\times0.2\arcsec$ 
slit and the G140L, G470L and G750L gratings, providing coverage over the wavelength ranges 1000\AA\ - 1700\AA\ and 3000\AA\ - 10000\AA\ 
in the observed frame. Spatial resolutions are 0.024 arcseconds pixel$^{-1}$ for the UV spectra, and 0.05 arcseconds pixel$^{-1}$ 
for the optical spectra. Spectral resolutions are a FWHM of 1.5 pixels at 1500\AA\ for the G140L grating, 1.3 pixels at 3200\AA\ for 
the G430L grating, and 1.6 pixels at 7000\AA\ for the G750L grating.  

\begin{figure*}
\begin{minipage}{170mm}
\epsfig{figure=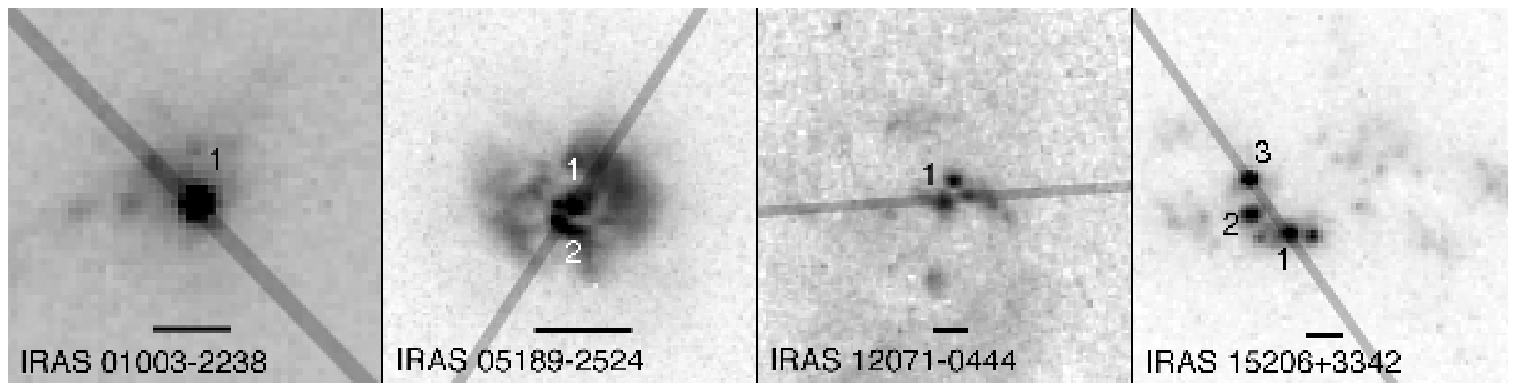,width=170mm}
\end{minipage}
\caption{HST WFPC2 images of our sample, taken from \citet{sur}, with the slit positions 
superimposed and the nuclear regions marked. North is up and east is to the left. The scale bar denotes 1Kpc.  
 \label{slits}}
\end{figure*}

In all cases the slit was centered on the putative nucleus whose luminosity and color were consistent with an obscured AGN \citep{sur}, 
and oriented so it also covered one or more off-nuclear `knots', if such knots were present. In Figure \ref{slits} we present the 
WFPC2 images of our sample, with the STIS slit positions superimposed. In Table \ref{knots} we list the magnitudes and radii of 
these knots, together with the stellar masses inferred by \citet{sur} under the assumption that all of the B and I band flux is stellar 
in origin. The total exposure times were 5500s for the G140L spectra, 2700s for the G470L spectra, and 1750s for the G750L spectra. 
Each observation was split into three or more exposures dithered with integer pixel shifts to allow the removal of cosmic rays and 
bad pixels. 

Individual exposures were reduced using the automated {\it calstis} pipeline at the STScI and the latest calibration files 
available as of January 2004. The exposures were then combined into a single image using the Image Reduction and 
Analysis Facility (IRAF) task {\it ocrreject}. Spectra were extracted using apertures corresponding to the same
physical size in the UV and optical for each object. For F01003-2238, F12071-0444 and F15206+3342, apertures of width 3 pixels for the optical 
spectra and 7 pixels for the UV spectra were used, corresponding to a size of approximately 350pc. For F05189-2524, where 
the two knots covered by the STIS slit lie $\sim300$pc from each other, apertures of width 5 pixels for the optical spectra and 11 
pixels for the UV spectrum were used, corresponding to a smaller physical size of approximately 220pc.

Line profiles and errors were measured using the {\it specfit} package within IRAF. This package can fit a wide variety of emission 
profiles, absorption profiles and continuum shapes to an input spectrum, and can disentangle multiple blended lines. The fitting itself 
was carried out via $\chi^{2}$ minimization, using a combination of Levenberg-Markwardt and Simplex algorithms. Input parameters for 
{\it specfit} were determined using the IRAF task {\it splot}. 

To deredden the UV spectra the color excess of the ionized gas, $E(B-V)_{gas}$ was derived from the observed and (assumed) intrinsic 
Balmer decrements:

\begin{equation}
E(B-V)_{gas} = 0.935 \times ln\frac{(H\alpha/H\beta)_{obs}}{(H\alpha/H\beta)_{int}}
\end{equation}

\noindent where intrinsic Balmer decrements of 2.85 and 3.1 were assumed, corresponding to heating by stars and an AGN respectively. This 
color excess was then converted to the reddening of the stellar populations via \citep{cal}:

\begin{equation}
E(B-V)_{stars} = 0.44E(B-V)_{gas}
\end{equation}

\noindent The extinction at some wavelength $\lambda$, $A_{\lambda}$, can then be computed using the prescription given by \citet{lei2002}:

\begin{eqnarray}\label{uvextinct}
A_{\lambda} =  E(B-V)_{stars} \times \nonumber \\
\left( 5.472 + \frac{0.671}{\lambda} - \frac{9.218\times10^{-3}}{\lambda^{2}} + \frac{2.62\times10^{-3}}{\lambda^{3}} \right)
\end{eqnarray}

\noindent This formula is valid over the wavelength range $0.097< \lambda(\mu m)<0.18$. The optical spectra were 
dereddened using the prescription given by \citet{ccm}\footnote{This paper also presents a prescription for dereddening 
UV spectra at wavelengths as small as 0.1$\mu$m, however their far-UV segment is derived using limited data, and 
therefore we do not adopt their prescription}. 

Line Full Widths at Half Maximum (FWHM) were corrected for the effects of finite instrument resolution using the quadrature method, where:

\begin{eqnarray}
FWHM^{2}_{real}=FWHM^{2}_{obs}-FWHM^{2}_{ins} 
\end{eqnarray}

\noindent This method assumes that the real and instrument profiles are Gaussians, and therefore gives line widths that are 
systematically too high for profiles that are more peaky than Gaussians. As such peaky line profiles are known to be present 
in AGN \citep{vei91}, the discussion of line widths in this paper should be regarded with caution.

\section{Results} \label{results}
\begin{figure*}
\begin{minipage}{150mm}
\epsfig{figure=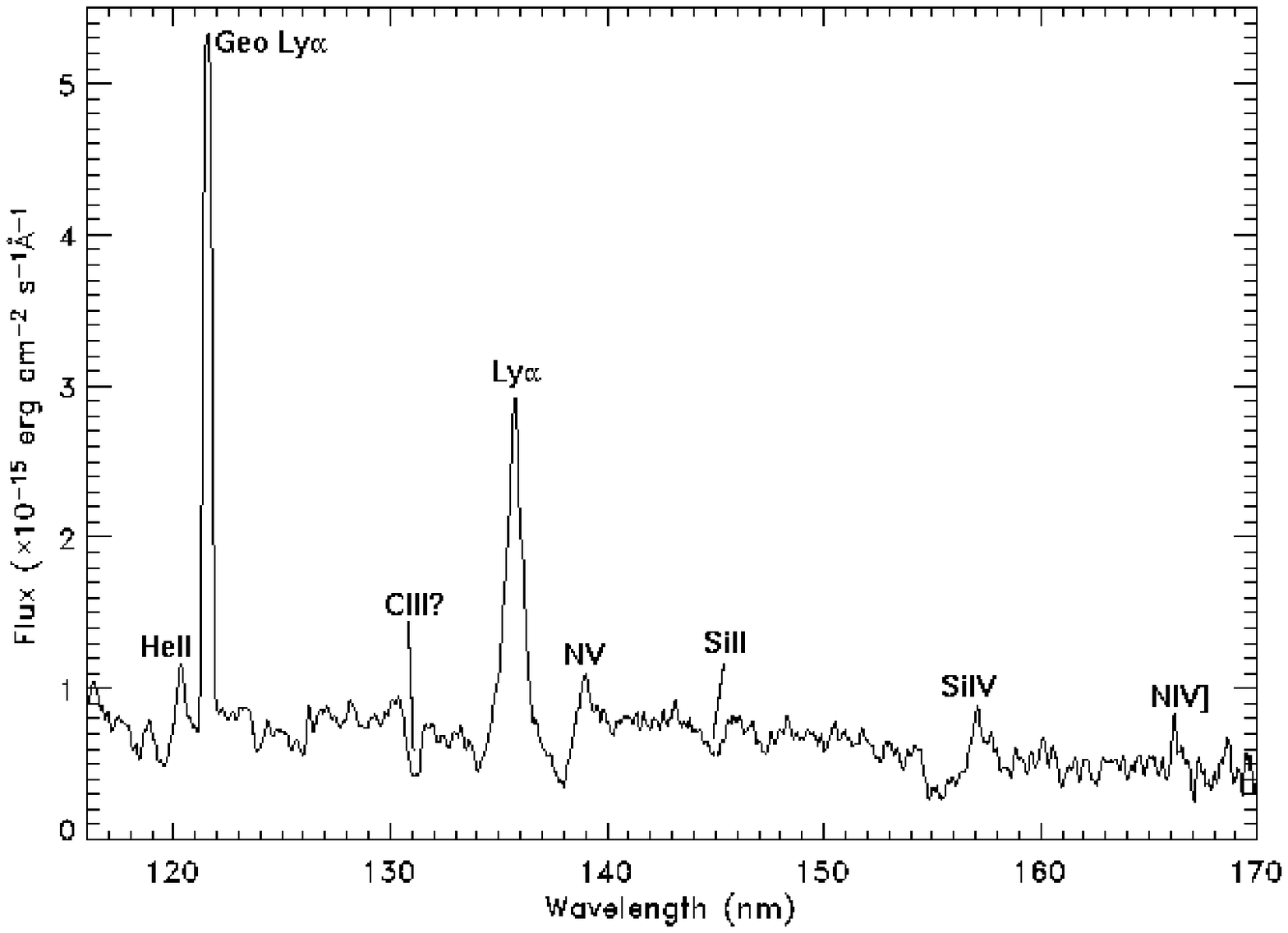,width=95mm}
\end{minipage}
\begin{minipage}{150mm}
\epsfig{figure=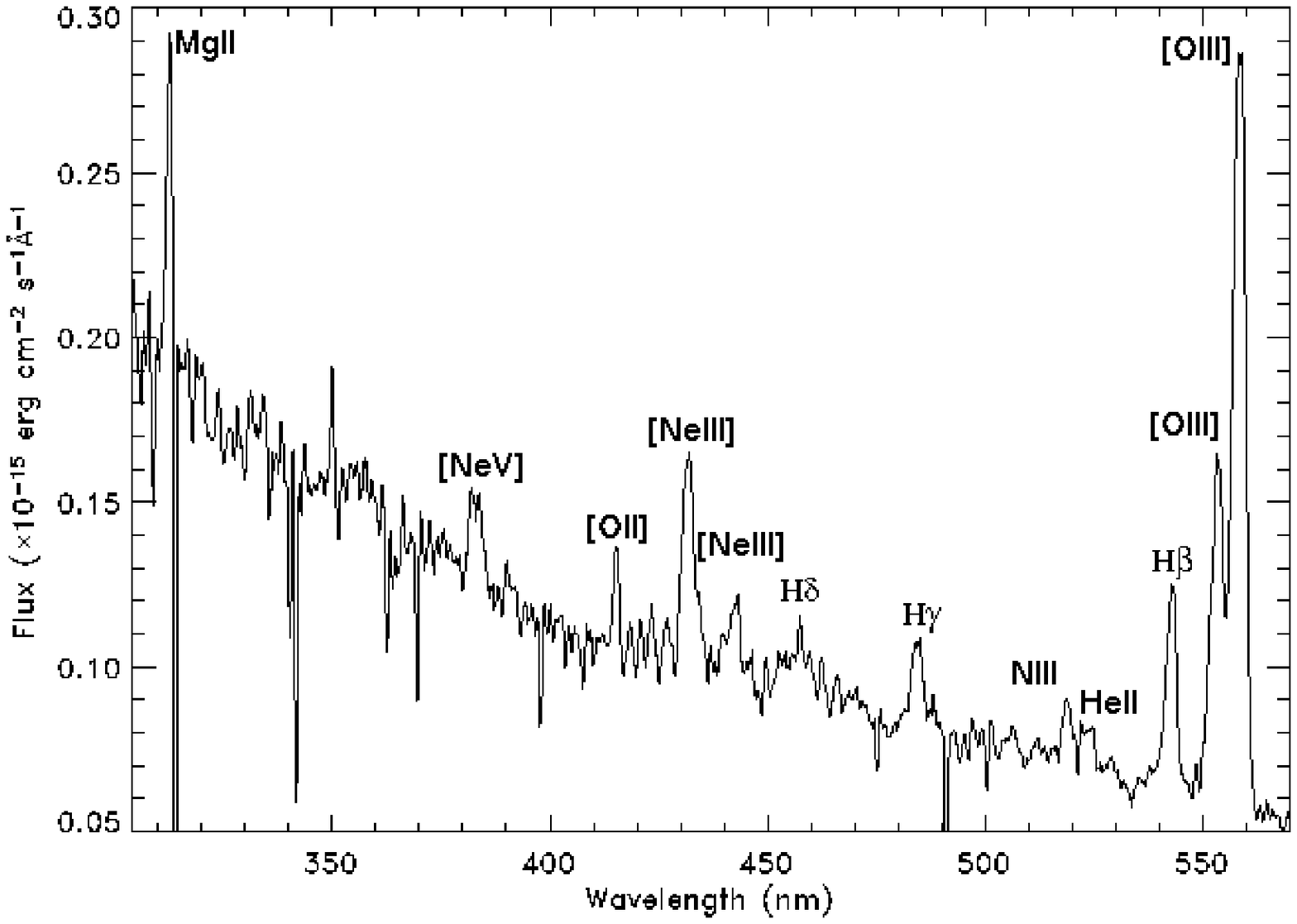,width=95mm}
\end{minipage}
\begin{minipage}{150mm}
\epsfig{figure=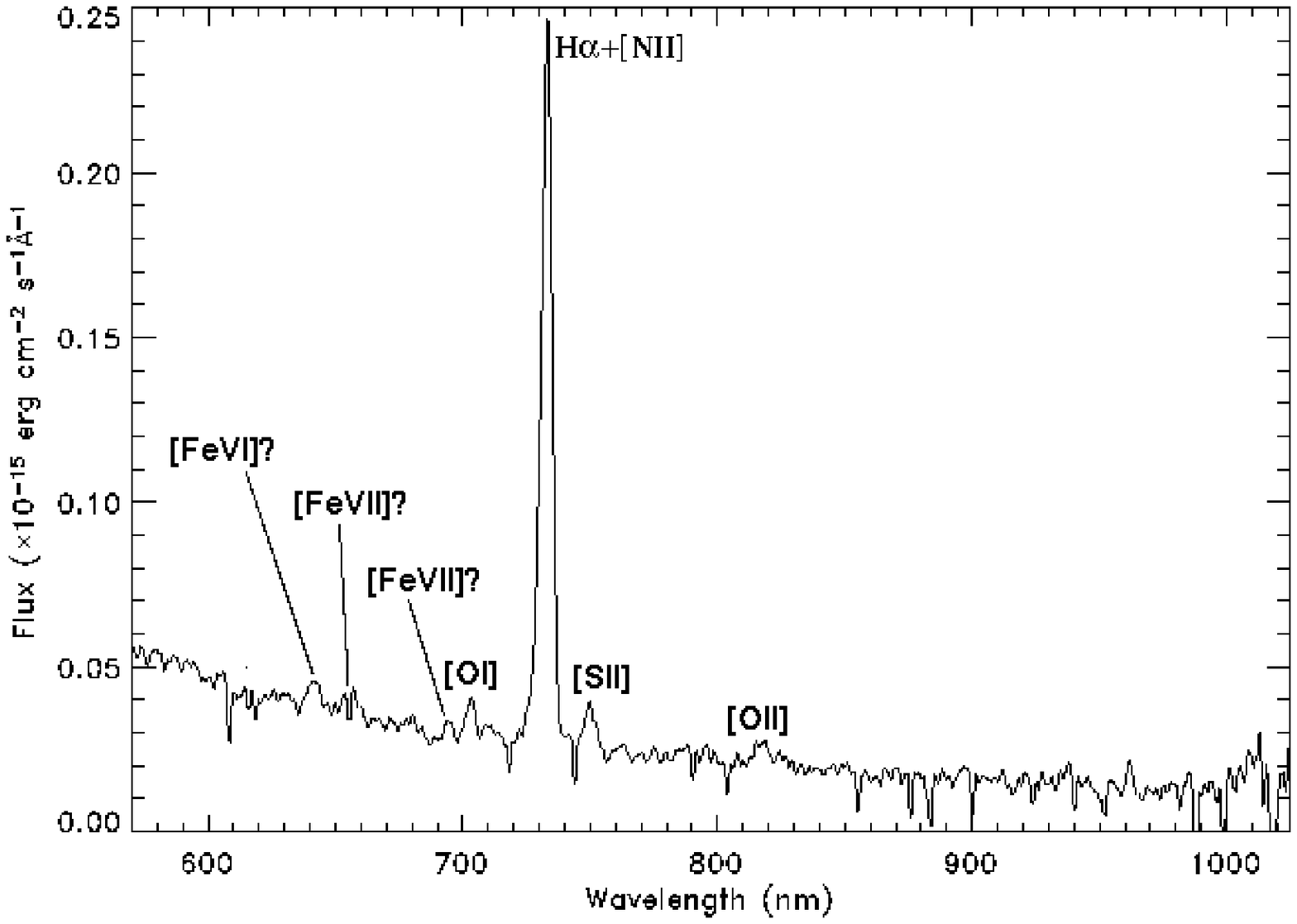,width=95mm}
\end{minipage}
\caption{G140L, G430L and G750L spectra of IRAS F01003-2238. The spectra have been smoothed with a 3 pixel boxcar 
and are plotted in the observed frame. Line IDs for the iron features at (observed-frame) 6416\AA, 6535\AA\ and 6941\AA\ 
are uncertain, see \S\ref{01003prop} for details.  \label{01003_spec}}
\end{figure*}

\begin{figure*}
\begin{minipage}{150mm}
\epsfig{figure=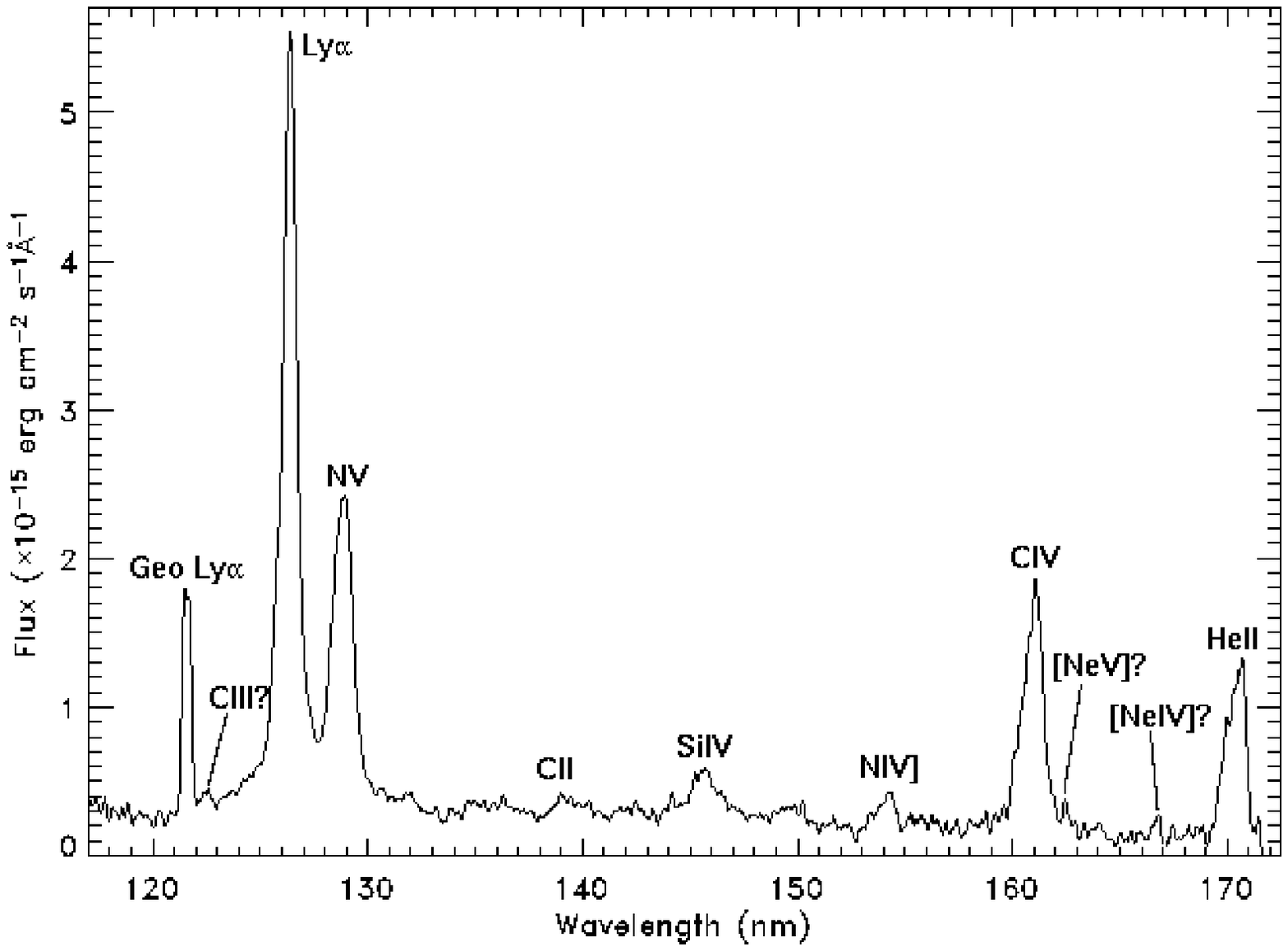,width=95mm}
\end{minipage}
\begin{minipage}{150mm}
\epsfig{figure=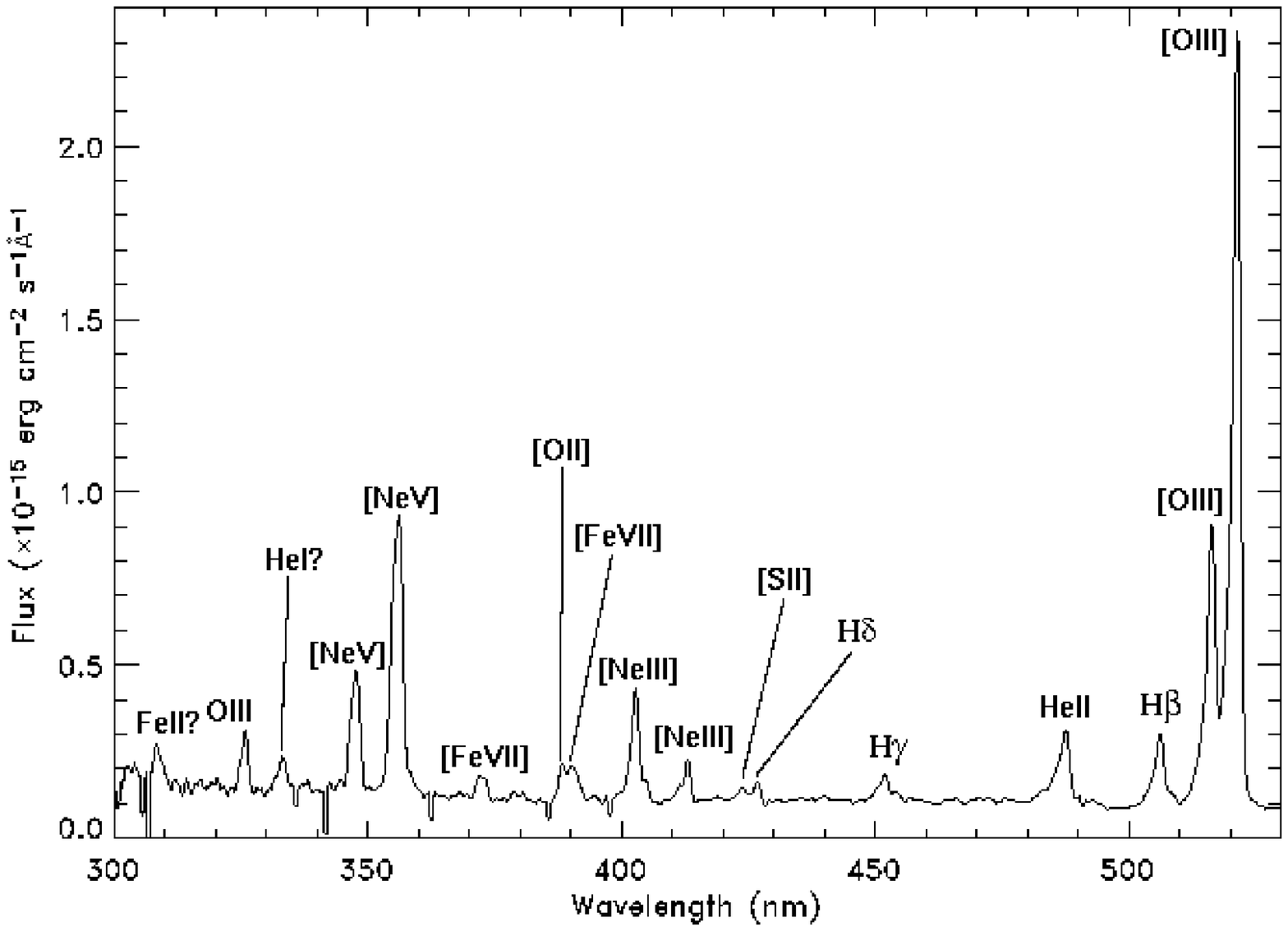,width=95mm}
\end{minipage}
\begin{minipage}{150mm}
\epsfig{figure=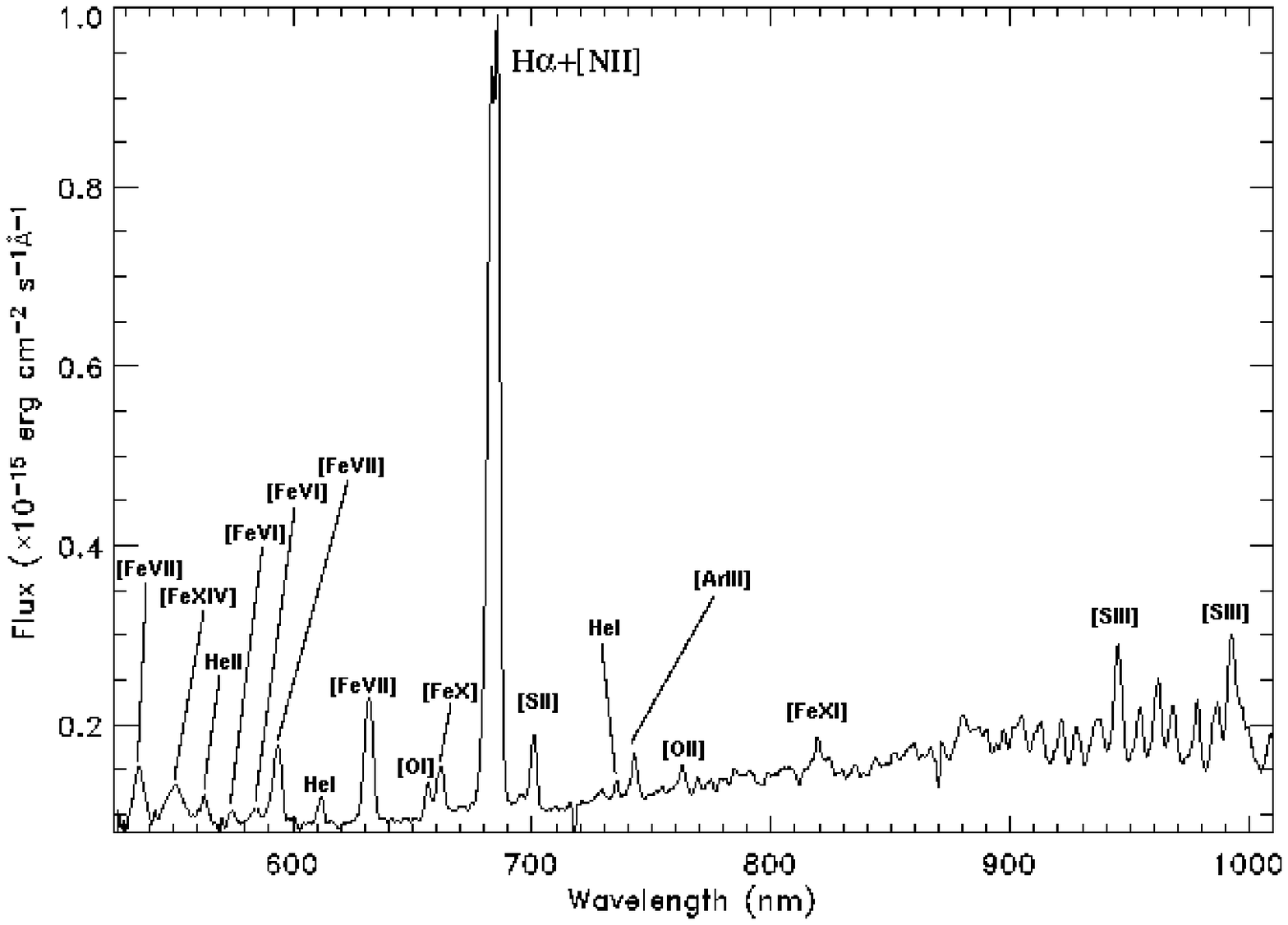,width=95mm}
\end{minipage}
\caption{G140L, G430L and G750L spectra of IRAS F05189-2524 knot 1. The spectra have been smoothed with a 3 pixel boxcar 
and are plotted in the observed frame. 
 \label{05189_spec}}
\end{figure*}

\begin{figure*}
\begin{minipage}{150mm}
\epsfig{figure=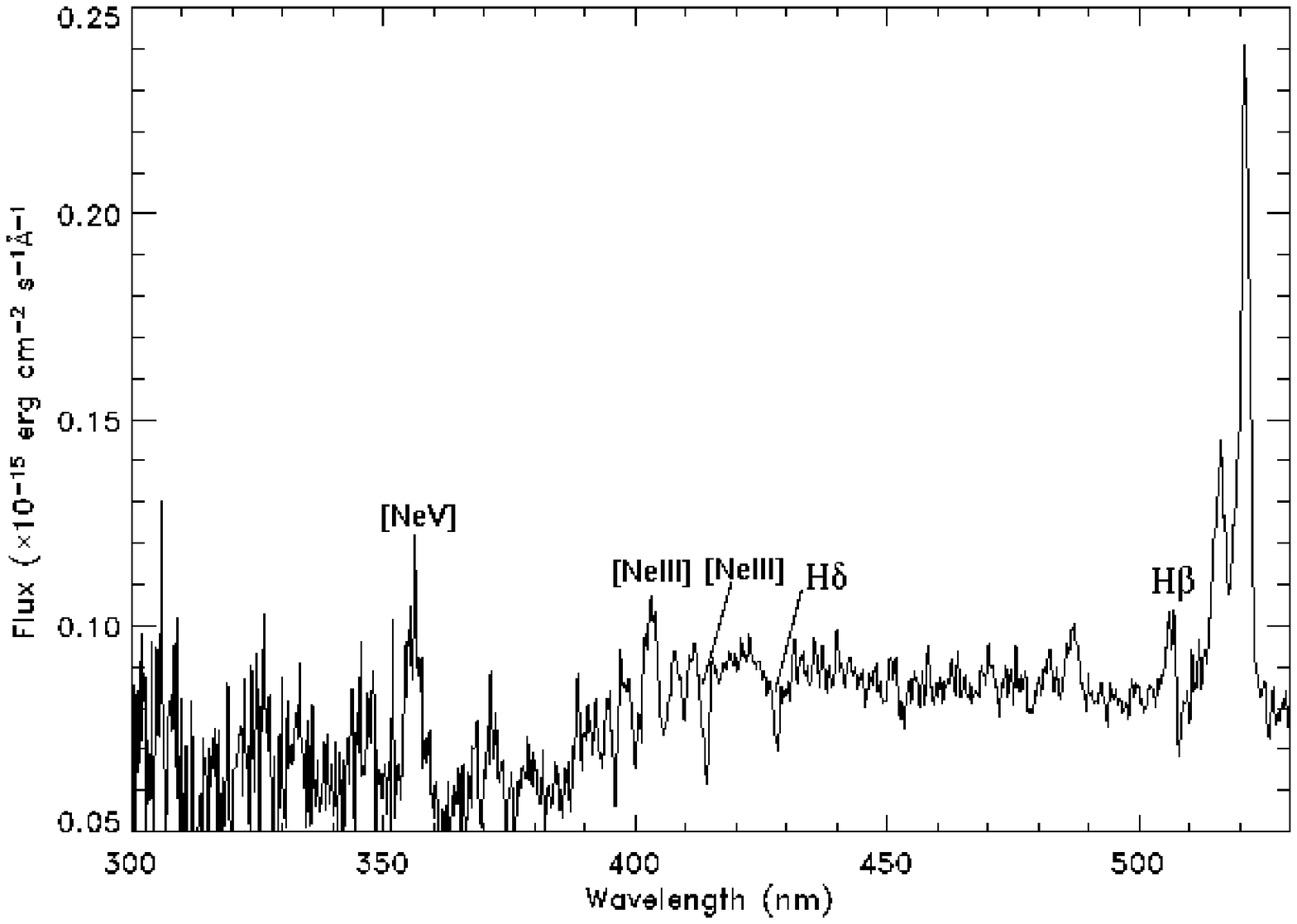,width=95mm}
\end{minipage}
\begin{minipage}{150mm}
\epsfig{figure=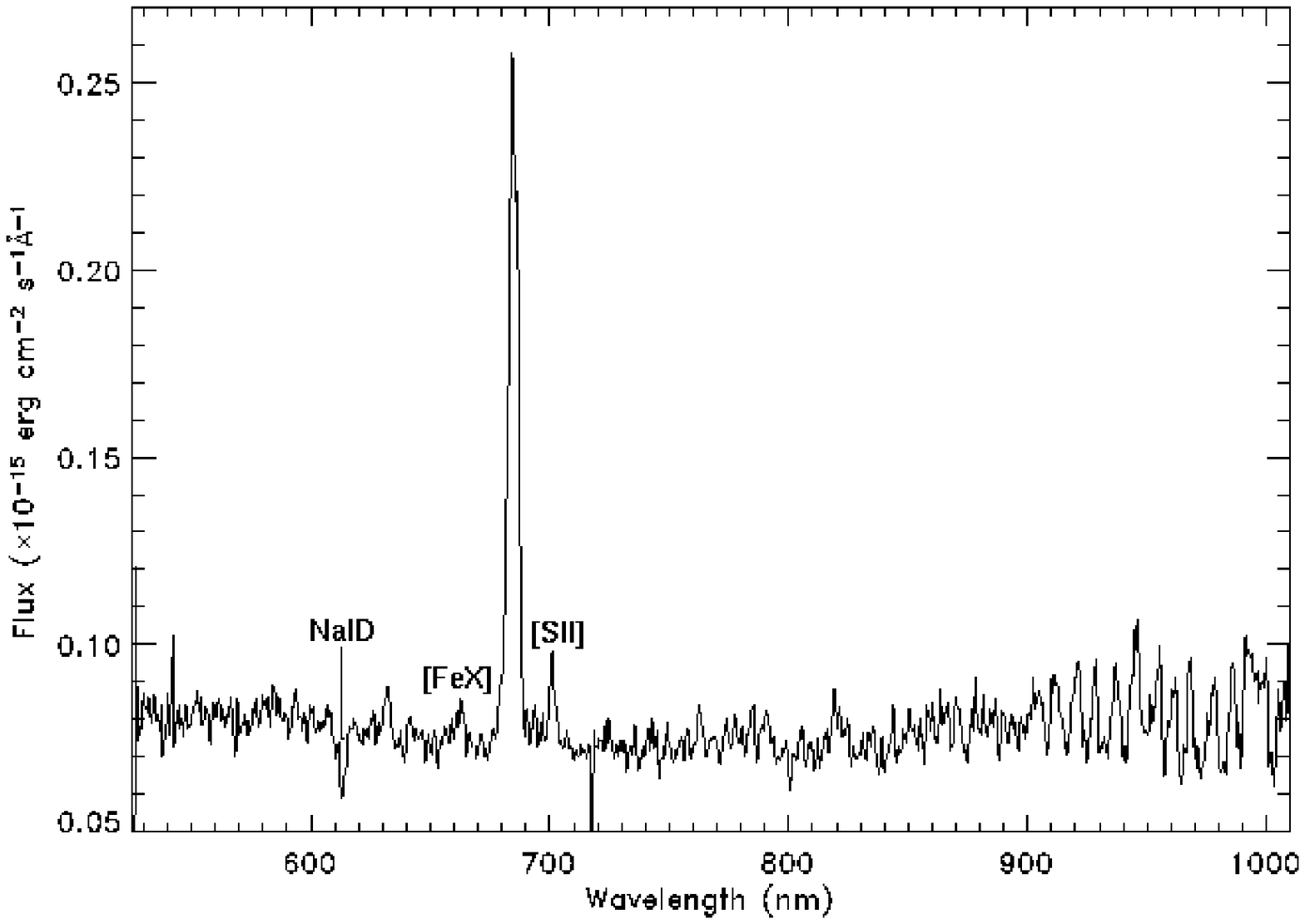,width=95mm}
\end{minipage}
\caption{G430L and G750L spectra of IRAS F05189-2524 knot 2. The spectra were extracted using a 1 pixel aperture to minimize 
contamination from the brighter knot, and {\it only those lines with no contamination from knot 1 are marked}. The spectra have been smoothed 
with a 3 pixel boxcar and are plotted in the observed frame. 
 \label{05189_2_spec}}
\end{figure*}

\begin{figure*}
\begin{minipage}{150mm}
\epsfig{figure=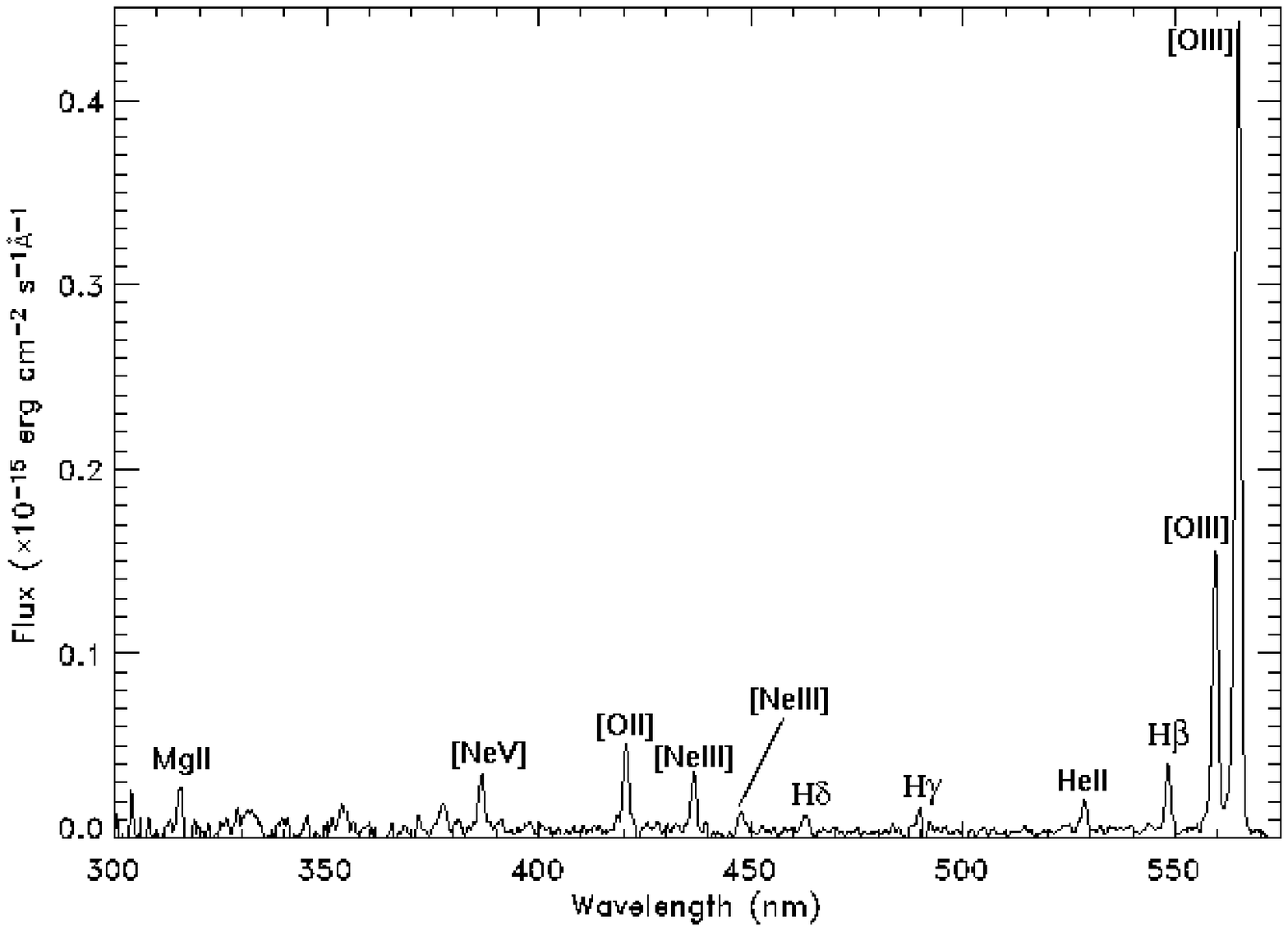,width=95mm}
\end{minipage}
\begin{minipage}{150mm}
\epsfig{figure=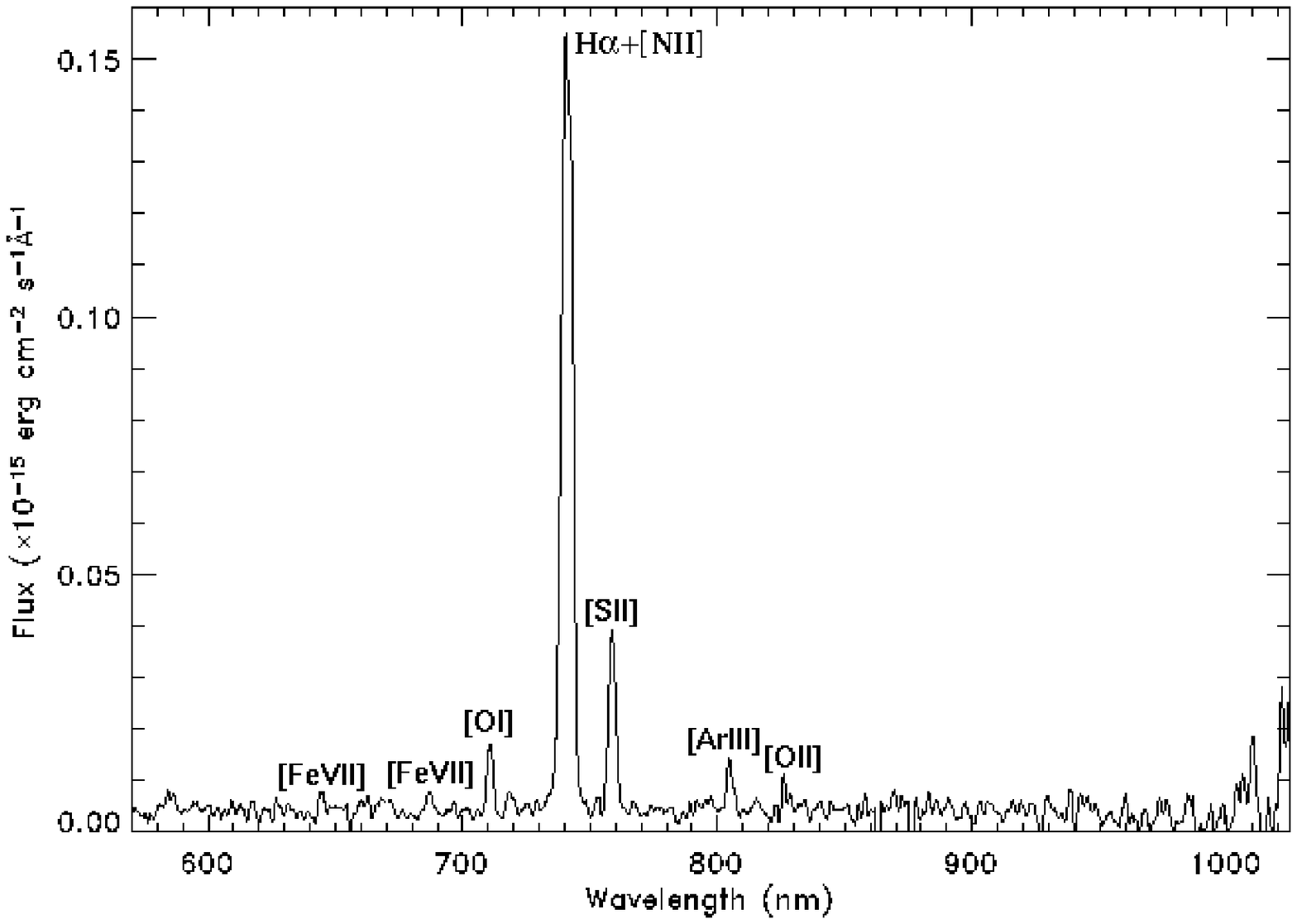,width=95mm}
\end{minipage}
\caption{G430L and G750L spectra spectra of IRAS F12071-0444. The spectra have been smoothed with a 3 pixel boxcar 
and are plotted in the observed frame. 
 \label{12071_spec}}
\end{figure*}

\begin{figure*}
\begin{minipage}{150mm}
\epsfig{figure=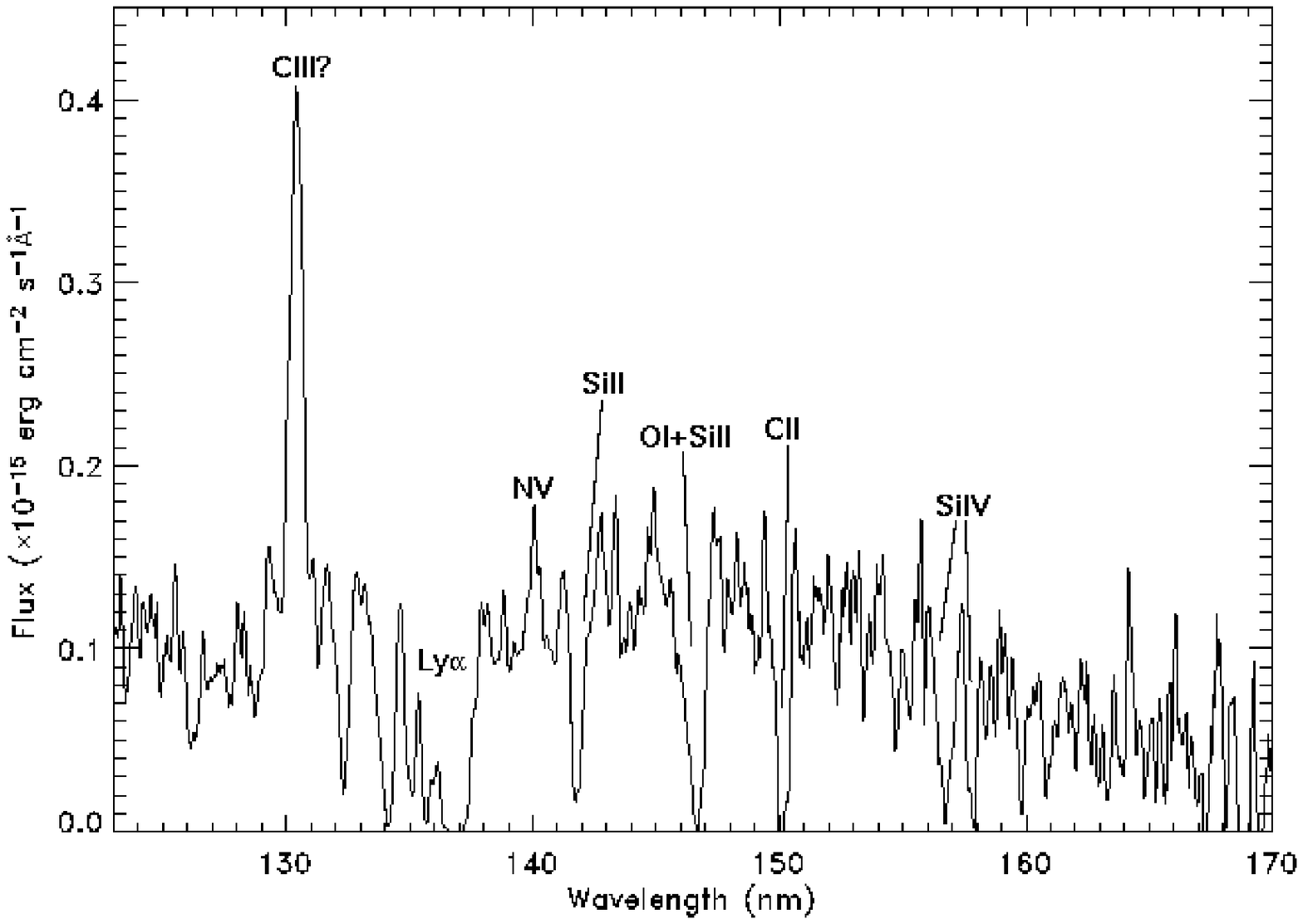,width=95mm}
\end{minipage}
\begin{minipage}{150mm}
\epsfig{figure=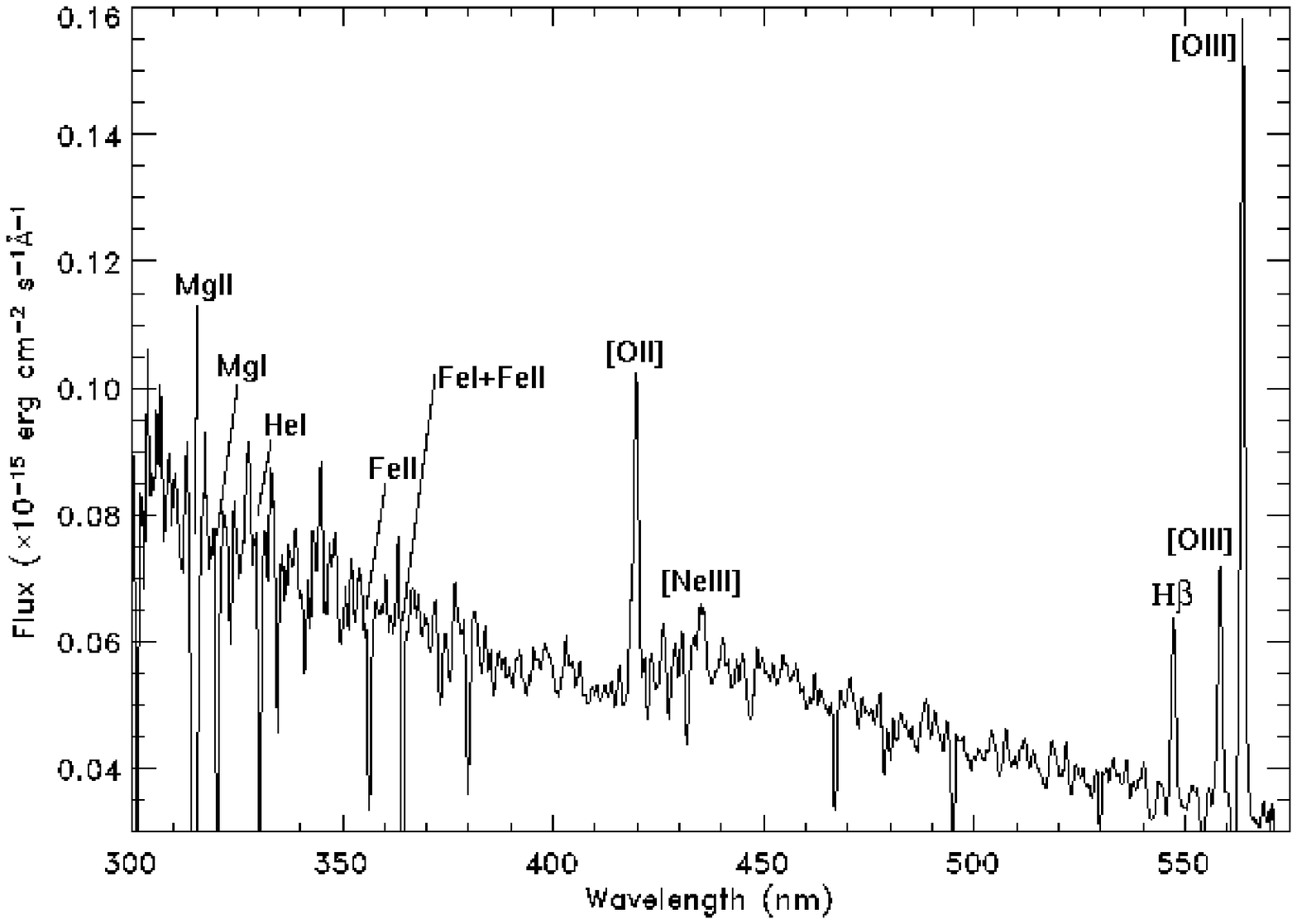,width=95mm}
\end{minipage}
\begin{minipage}{150mm}
\epsfig{figure=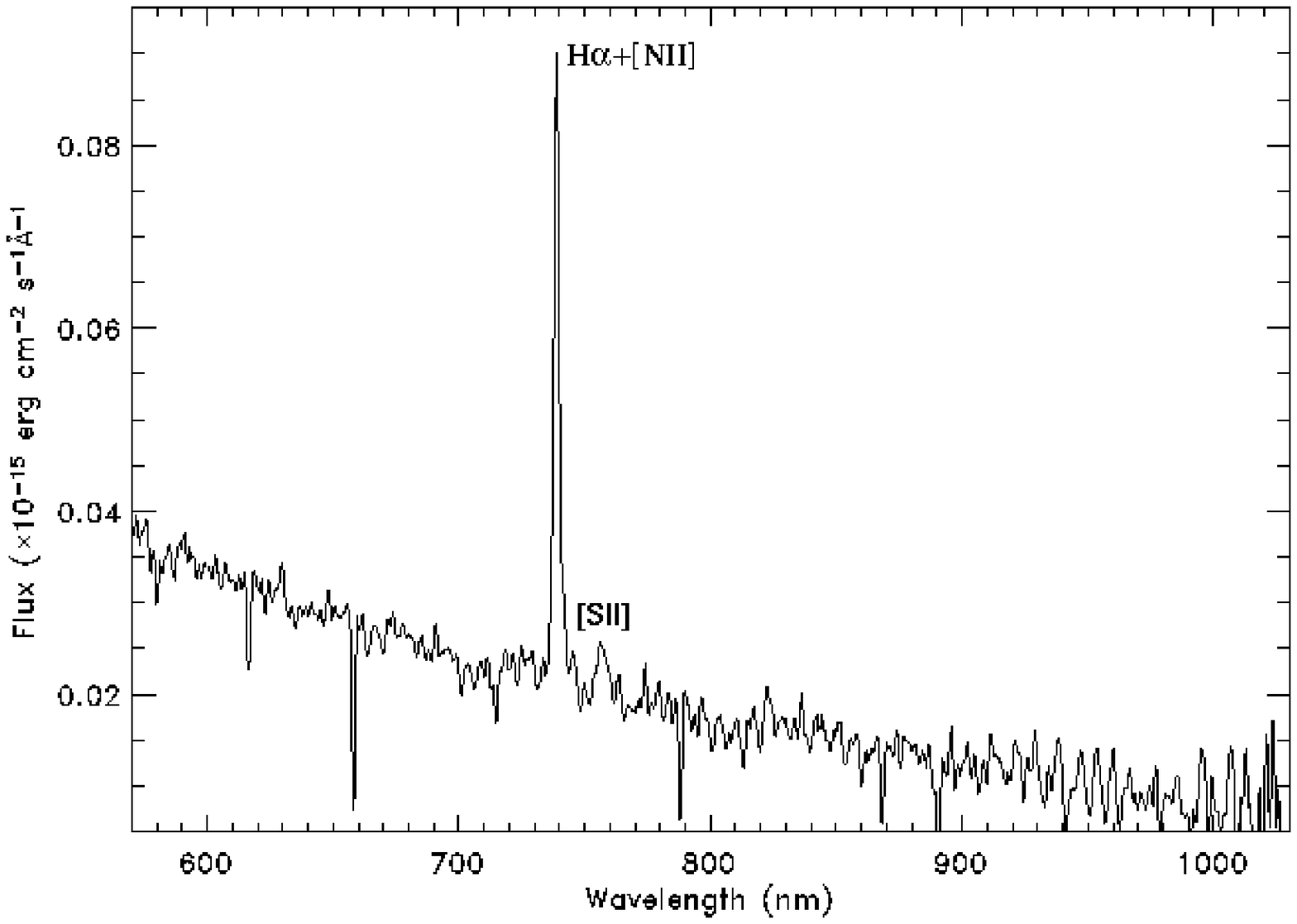,width=95mm}
\end{minipage}
\caption{G140L, G430L and G750L spectra of IRAS F15206+3342 knot 1. The spectra have been smoothed with a 3 pixel boxcar 
and are plotted in the observed frame. The narrow `absorption' features in the I band spectrum are cosmic ray residuals.
 \label{15206_587_spec}}
\end{figure*}

\begin{figure*}
\begin{minipage}{150mm}
\epsfig{figure=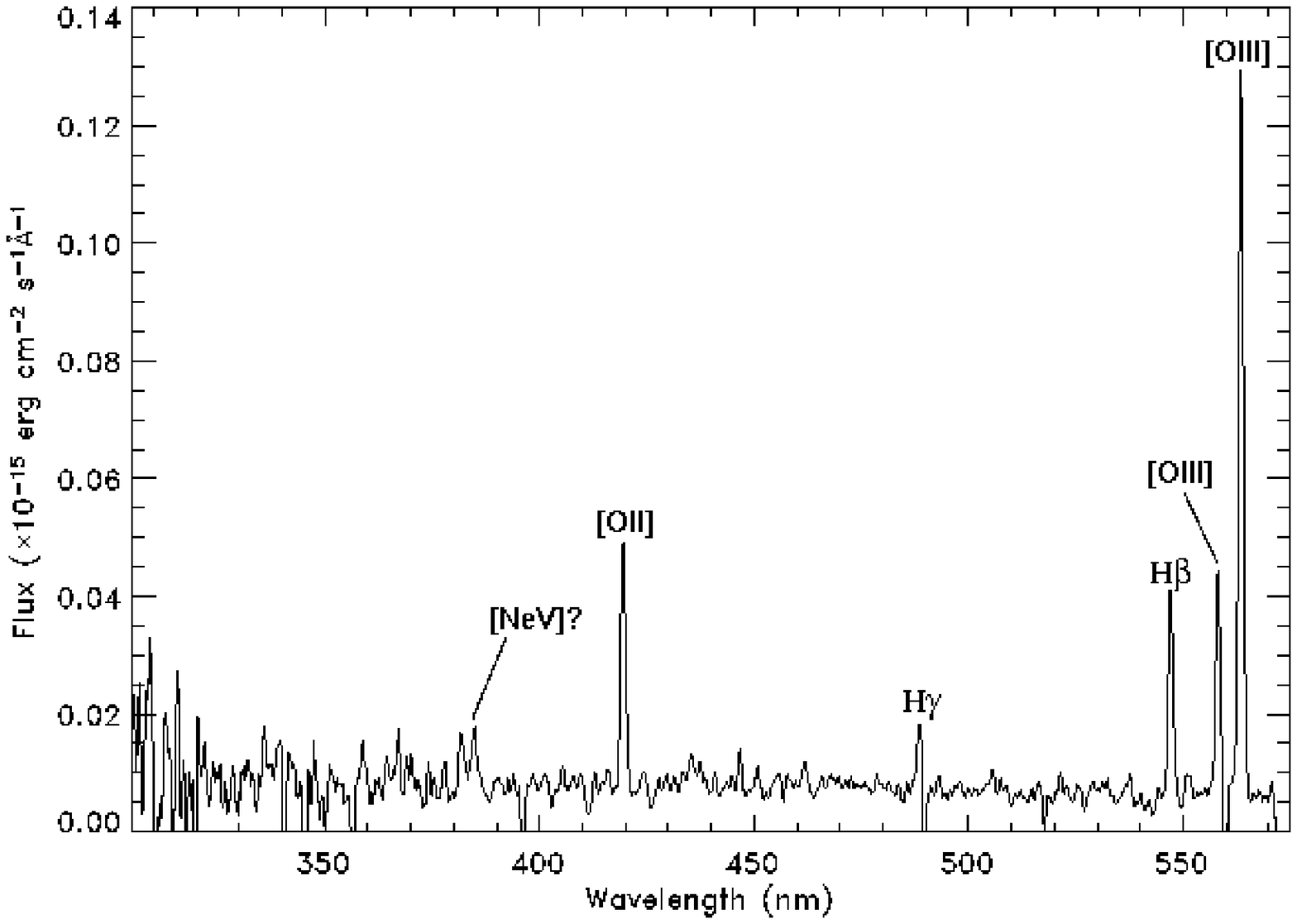,width=95mm}
\end{minipage}
\begin{minipage}{150mm}
\epsfig{figure=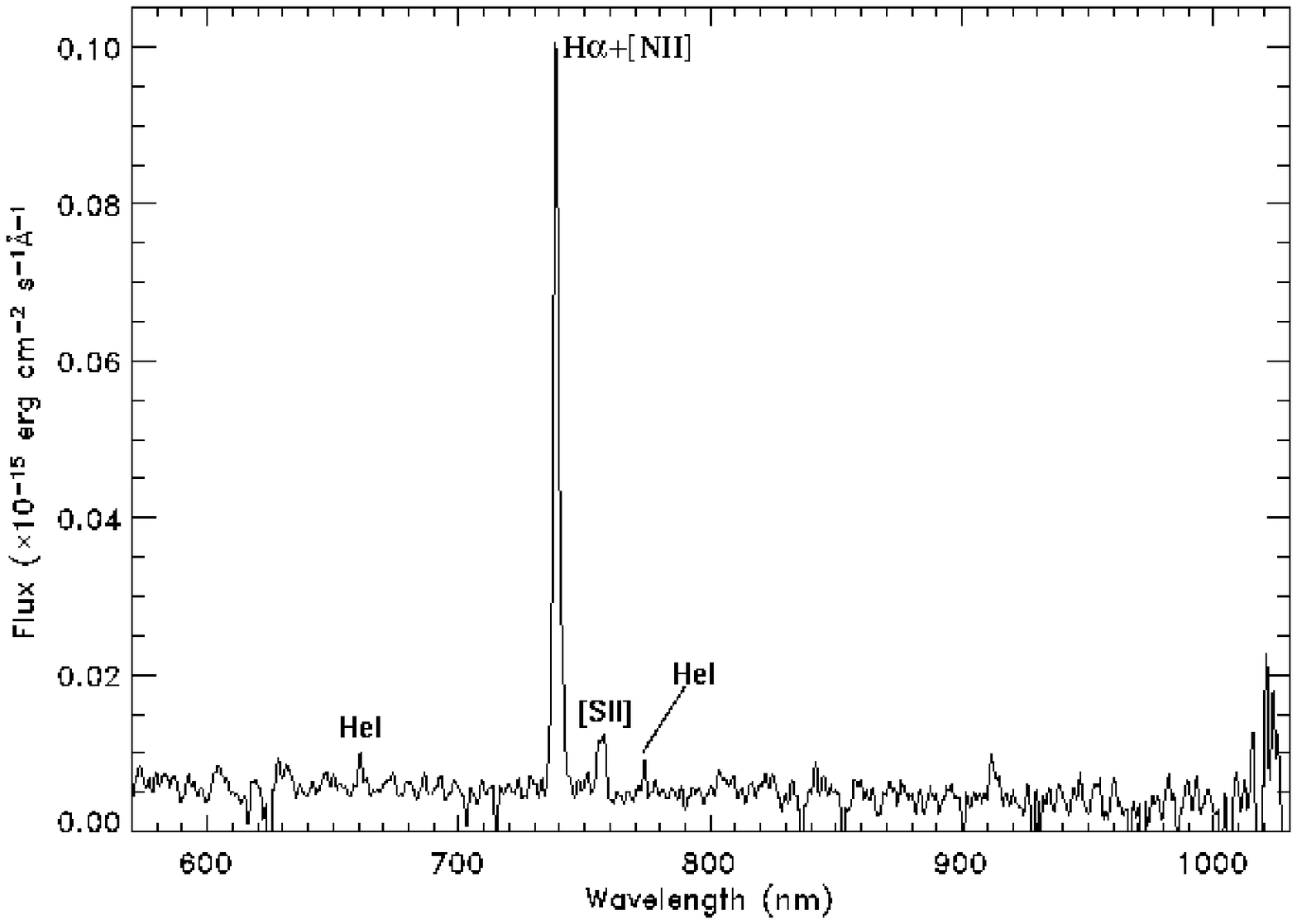,width=95mm}
\end{minipage}
\caption{G430L and G750L spectra of IRAS F15206+3342 knot 2. The spectra have been smoothed with a 3 pixel boxcar 
and are plotted in the observed frame. 
 \label{15206_594_spec}}
\end{figure*}

\begin{figure*}
\begin{minipage}{150mm}
\epsfig{figure=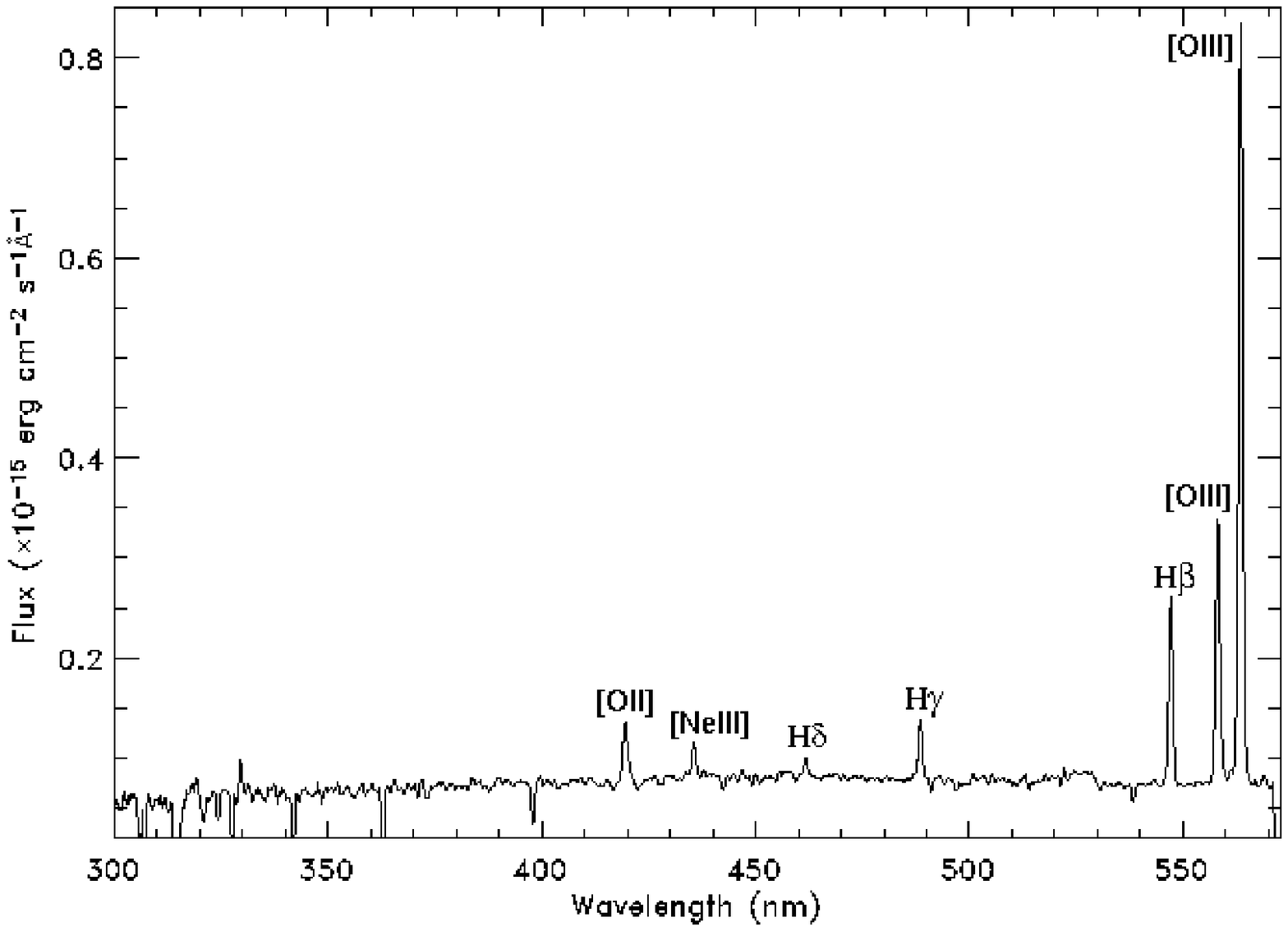,width=95mm}
\end{minipage}
\begin{minipage}{150mm}
\epsfig{figure=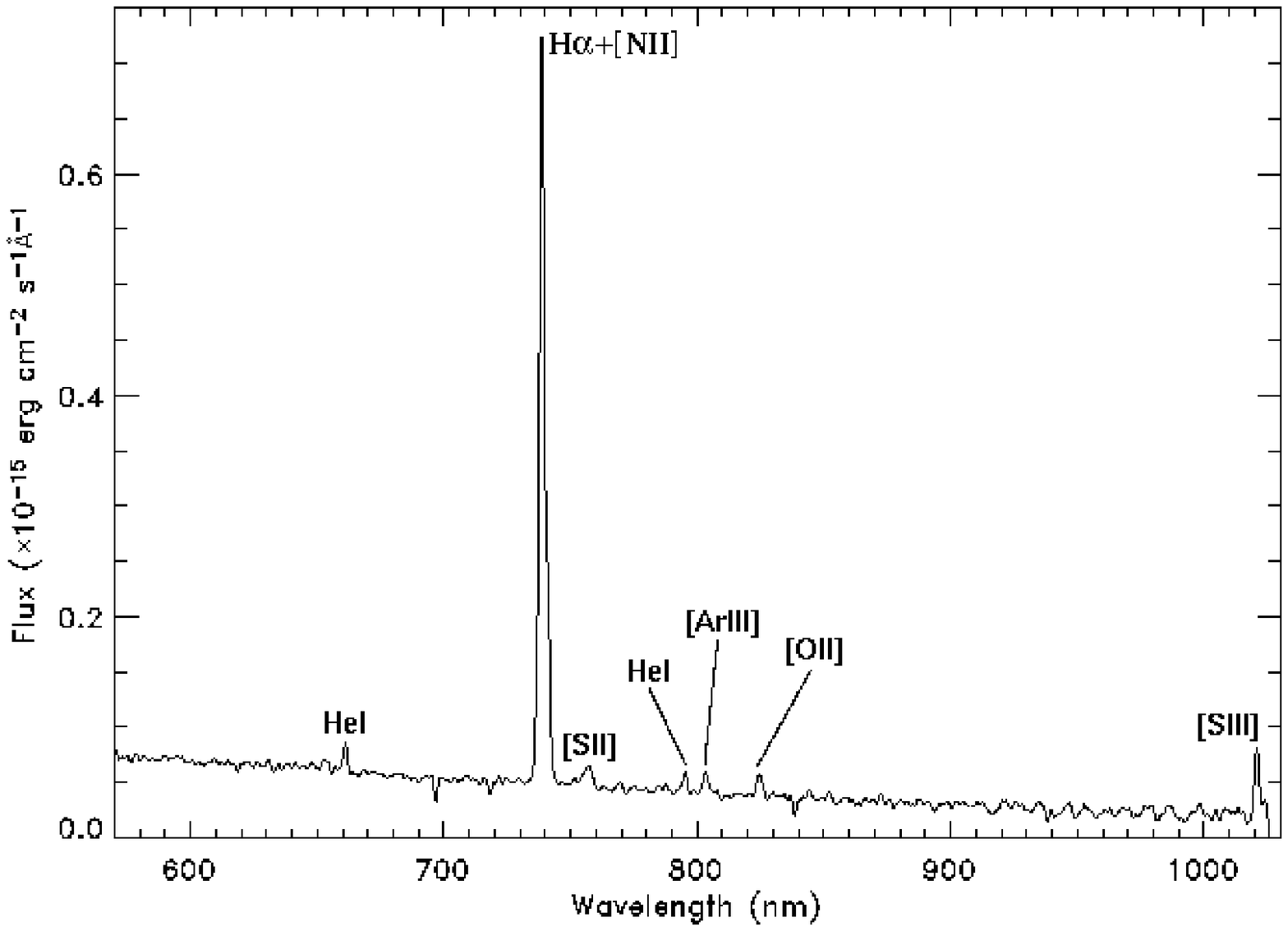,width=95mm}
\end{minipage}
\caption{G430L and G750L spectra of  IRAS F15206+3342 knot 3. The spectra have been smoothed with a 3 pixel boxcar 
and are plotted in the observed frame. 
 \label{15206_601_spec}}
\end{figure*}

The UV and optical spectra for each ULIRG are presented in Figures \ref{01003_spec}, 
\ref{05189_spec}, \ref{05189_2_spec}, \ref{12071_spec}, \ref{15206_587_spec}, \ref{15206_594_spec} \& \ref{15206_601_spec}. 
The UV and optical emission line fluxes are presented in Tables \ref{ulinefluxes} \& \ref{olinefluxes} respectively. Widths of 
selected emission lines are given in Table \ref{linewidths}. 

\begin{figure*}
\begin{minipage}{150mm}
\epsfig{figure=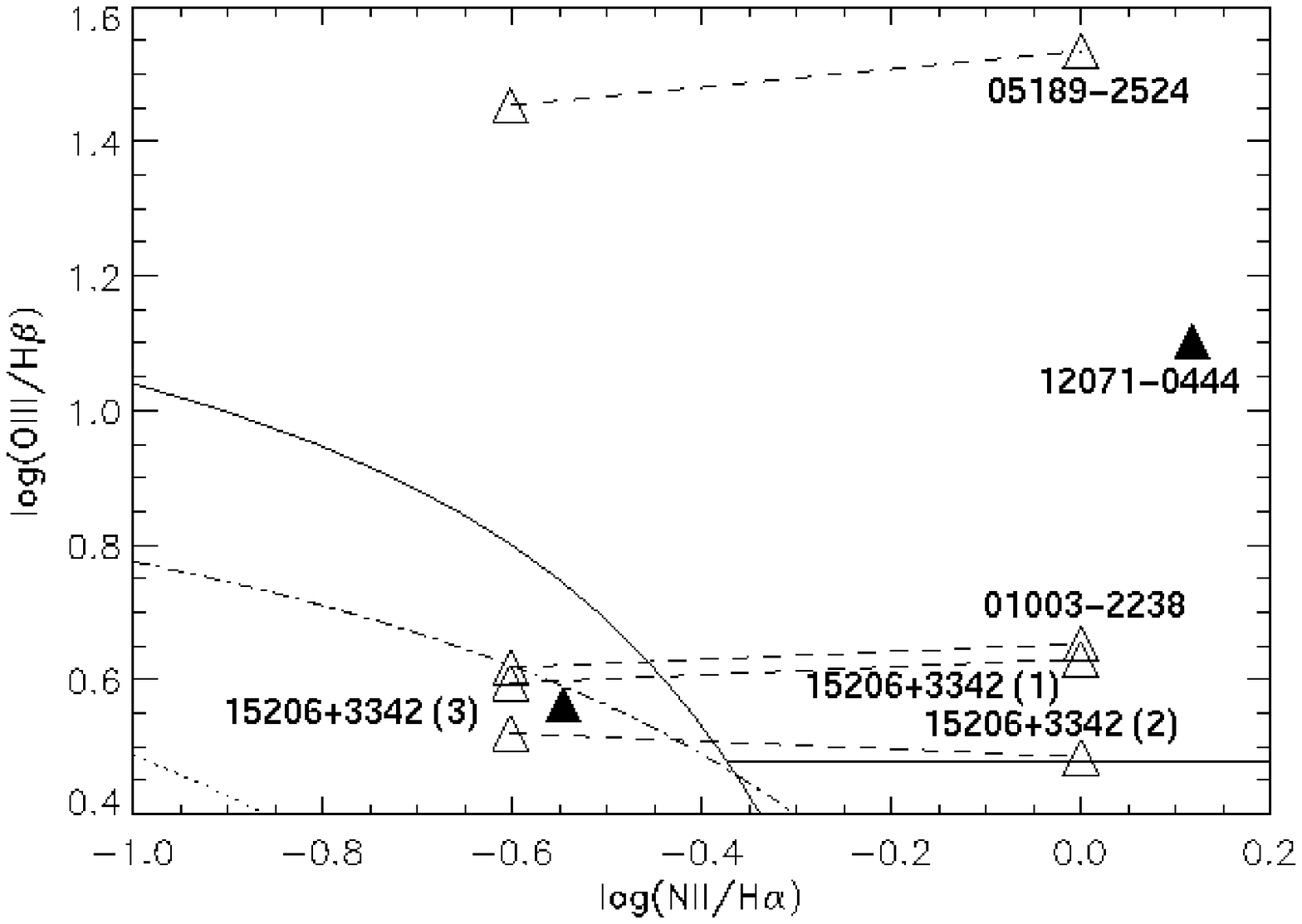,width=90mm}
\end{minipage}
\begin{minipage}{150mm}
\epsfig{figure=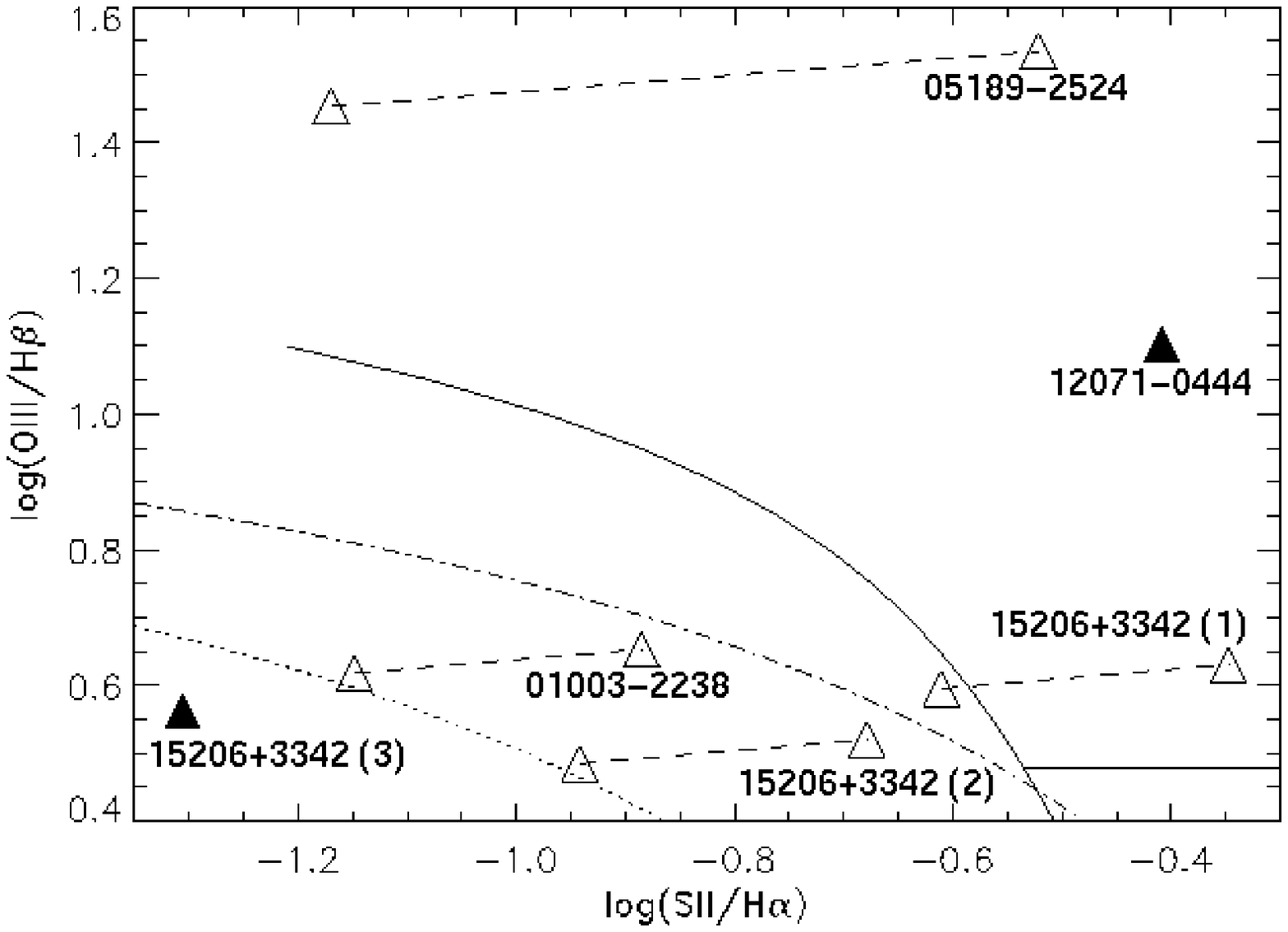,width=90mm}
\end{minipage}
\begin{minipage}{150mm}
\epsfig{figure=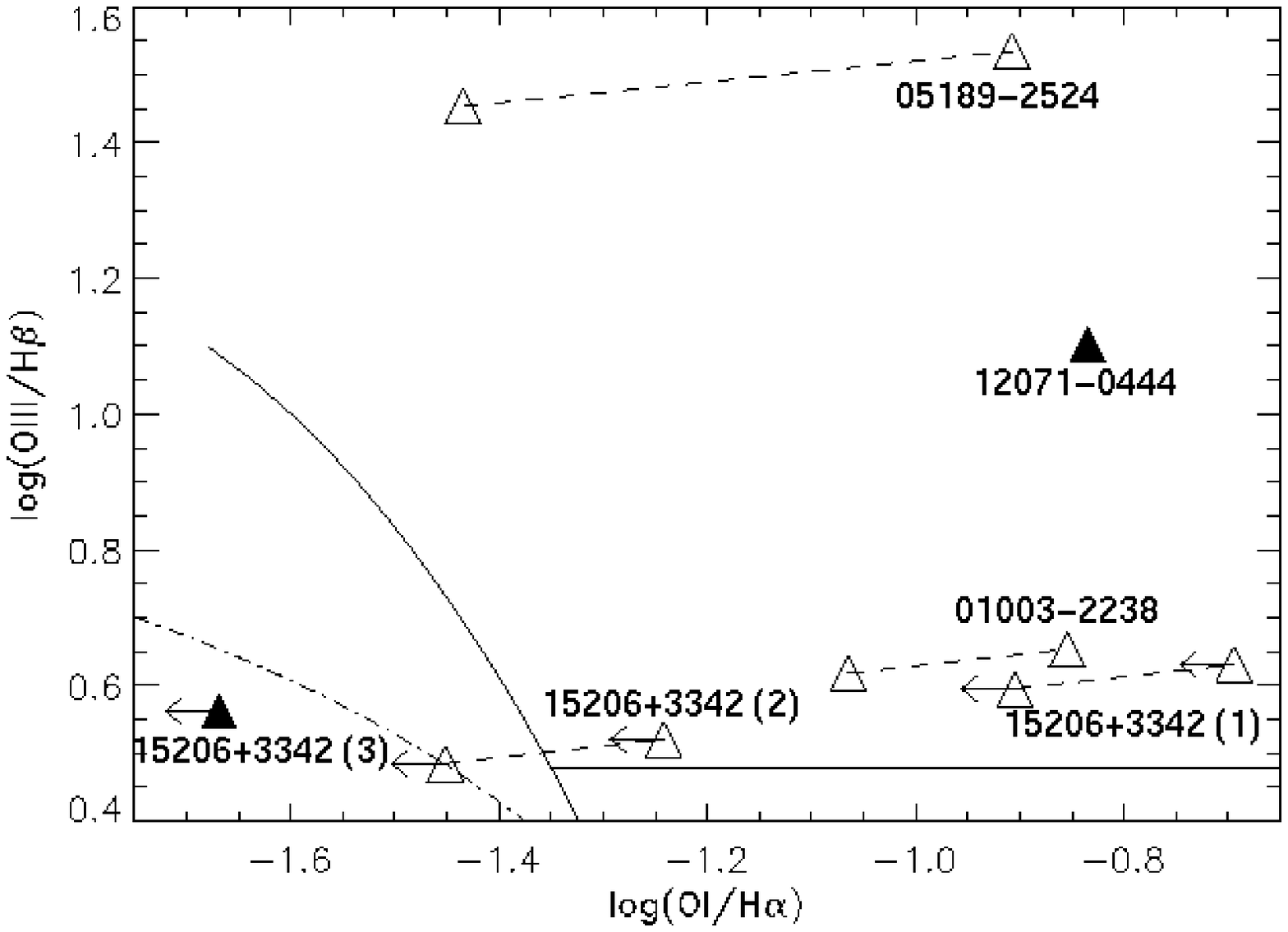,width=90mm}
\end{minipage}
\caption{ Line ratio diagnostic diagrams. All fluxes used in these ratios have been corrected for the effects of extinction using 
the methods described in \S \ref{obsanl}.  Filled symbols show ratios where all lines have measured values, outline 
symbols connected by a line represent those objects where H$\alpha$ and [NII] could not be resolved and where two 
intrinsic values for the Balmer decrement (3.1 and 2.85) were therefore assumed. Arrows indicate upper limits. The curved 
solid lines are taken from \citet{vo87}. The horizontal solid lines demarcate Sy2's from LINERS \citep{hek}. 
Dotted lines are upper limits from \citet{dop} for zero age instantaneous bursts, dot-dashed lines are upper limits for 
continuous bursts from \citet{kewa}.  \label{linediags}}
\end{figure*}

All of the spectra show at least two emission lines, while a few also show absorption lines. These lines are a powerful tool in determining 
physical conditions in the ionized gas in our sample. In the optical, diagnostics using the stronger emission lines can be used to constrain 
the temperature, density and kinematics of the hot gas in galaxies. In this paper we have used the three line ratio diagnostics presented by 
\citet{vo87} as the primary method to distinguish between HII and AGN spectra;
[OIII] $\lambda 5007 / $H$\beta$ vs. [NII] $\lambda 6583 / $H$\alpha$, 
[OIII] $\lambda 5007 / $H$\beta$ vs. [SII] $\lambda\lambda 6716,6731/$H$\alpha$, and
[OIII] $\lambda 5007 / $H$\beta$ vs. [OI] $\lambda 6300 / $H$\alpha$.
For those spectra classified as AGN we adopt the definition that a LINER is an AGN with log([OIII] $\lambda 5007 / $H$\beta) < 0.477$. For those 
spectra classified as HII regions we have used the theoretical upper limits on these line ratio diagnostics for both continuous star formation 
\citep{kewa} and instantaneous bursts \citep{dop}. All of these diagnostics are presented in Table \ref{lineratios}, and plotted in Figure \ref{linediags}. 
We supplemented these with two additional diagnostics; [OIII] $\lambda 5007 / $H$\beta$ vs. [OII] $\lambda 3727$/[OIII] $\lambda 5007$  \citep{bpt} and 
[OIII] $\lambda 5007 / $H$\beta$ vs. [OII] $\lambda\lambda 7320,7330/$H$\alpha$ \citep{otv}. 

Spectroscopy in the ultraviolet, despite suffering significant extinction uncertainties, is the best way of directly studying the hottest, 
youngest stars, as at wavelengths longward of 3000\AA\ features from these stars are weak, and are often blended with nebular emission lines. 
The shapes of the stronger UV lines can be used to distinguish between starburst and AGN activity and shocks vs. photoionization, and determine 
parameters for the stellar initial mass function (IMF). Here we have analysed the UV spectra of our sample using both UV line ratio diagnostics, 
and modelling using the spectral synthesis package {\it Starburst99} v4.0 \citep{lei1999}. 

In the optical, star formation rates are commonly estimated using the fluxes of recombination lines, particularly H$\alpha$, calibrated using spectral 
synthesis models (see \citet{ken} for a review). Though the scatter in the calibration can be up to 30$\%$, due to differences between spectral 
synthesis models, this method has the advantages of high sensitivity, and direct coupling between the H$\alpha$ flux and the (massive) star formation 
rate. The primary disadvantage, for the purposes of our study, is that the H$\alpha$ flux from stars must be disentangled from the nebular 
[NII]$\lambda\lambda 6548,6583$ lines, and from any broad H$\alpha$ line arising from an AGN. In most cases it was not possible to reliably isolate the 
H$\alpha$ flux arising from star formation (with errors of a factor of two or more, see \S\S\ref{01003} - \ref{15206} for more details); we have 
therefore, for consistency, estimated optical star formation rates for all objects using the extinction corrected luminosity of the [OII] $\lambda 3727$ 
line, and the prescription presented by \citet{ken}:

\begin{equation}
SFR = 1.4\pm0.4\times10^{-41}L([OII]) \label{oiisfr}
\end{equation}

\noindent where the SFR is in units of $M_{\odot}$ yr$^{-1}$ and L[OII] is in units of ergs s$^{-1}$. The luminosity of [OII] $\lambda 3727$ is 
sensitive to the metallicity and ionization state of the gas in the narrow line region, and the calibration for [OII] $\lambda 3727$ as a star 
formation rate indicator is derived empirically through relationships derived for H$\alpha$. The errors on the star formation rates from 
Equation \ref{oiisfr} are therefore {\it at least} 30\%, and these star formation rates should therefore be regarded with caution. We also emphasize 
that, for those `knots' that contain an AGN, the star formation rates derived using Equation \ref{oiisfr} should be regarded as upper limits.

Where possible, we estimated electron densities and temperatures using the IRAF task {\it temden}. The commonly used line ratios for estimating 
electron temperatures are [OIII] $\lambda\lambda 4959,5007$ / [OIII] $\lambda 4363$ and [NII]$\lambda 5755$/[NII]$\lambda 6583$, however 
[OIII] $\lambda 4363$ and [NII]$\lambda 5755$ are not detected in any galaxy of our sample. Electron temperatures were therefore estimated using 
the line ratio [OII] $\lambda 3727$ / [OII] $\lambda 7325$. To estimate electron densities we used the [SII] $\lambda 6716$ / [SII] $\lambda 6731$ 
line ratio, though this was only possible in a few cases due to the blending of the two [SII] lines.

\subsection{IRAS F01003-2238} \label{01003}
\subsubsection{Background}
This object contains a single nucleus embedded in a small halo of extended emission, with a series of small blue knotlike features 
extending $1\arcsec$ to the north-west \citep{sur}. A serendipitous observation of this object by ROSAT did not detect any X-ray 
emission, and gave a $1\sigma$ upper limit to the soft X-ray luminosity of $2.14\times10^{40}$ ergs s$^{-1}$ \citep{sst}. Previous 
efforts to identify the power source in this galaxy have produced conflicting results. An early study by \citet{ahm}, using ground-based 
spectroscopy of the whole galaxy, identified several features characteristic of Wolf-Rayet stars (HeII $\lambda 4686$, HeI $\lambda 5876$, 
and tentatively NIII $\lambda\lambda\lambda 4634,4640,4642$ CIII $\lambda 5696$, NIV $\lambda 5737$, and several features around 7900\AA, 
though see also \citet{con}), and concluded that this galaxy contains approximately $10^{5}$ WN-type Wolf-Rayet stars, consistent with 
results from 2-D spectroscopy \citep{wca}. \citet{sur} however postulate that the central nucleus is an AGN, based on its position in 
optical colour-colour plots, and mid-infrared spectroscopy suggests that the IR emission is not dominated by a starburst \citep{tra}. 
Near-infrared spectroscopy \citep{vei97} however shows no evidence for a hidden broad line region.

\subsubsection{Emission \& absorption line properties}\label{01003prop}
The data show a single, bright spectrum in the UV and optical bands. In the UV spectrum Ly$\alpha$ emission is present, together with weaker 
emission lines of NV $\lambda 1240$, SiIV $\lambda\lambda 1394,1403$, and NIV] $\lambda 1488$. There are also a number of absorption systems, 
and weak, but distinct PCygni profiles on the SiIV doublet and NV lines, which imply winds from O supergiants. The SiIV doublet is commonly 
seen in O star photospheres, however the NIV] $\lambda 1488$ line is only seen in winds from W-R stars. We also clearly detect the Wolf-Rayet 
star signature triplet NIII$\lambda\lambda\lambda 4634,4640,4642$ (which may be blended with CIV $\lambda 4648$). We do not however  detect the 
`bump' between 7875\AA\ and 7965\AA\ claimed by \citet{ahm}. This could indicate either that this feature does not exist, or that it is produced 
outside the nuclear region. The B band spectrum contains the [NeV]$\lambda 3426$ line, which is thought to originate in AGN. The emission lines 
in the UV and optical spectra are spatially symmetric with respect to the continuum, with extents of 550-600pc for the Ly$\alpha$ and stronger 
optical lines, and $\sim280$pc for the high ionization UV lines (for F01003-2238 one pixel in the UV frame corresponds to $\sim$55pc and 1 pixel 
in the optical corresponds to $\sim$110pc). The line profiles of H$\beta$ and [OIII] are also symmetric, and show no blue wings. This is at odds 
with ground-based spectra \citep{ahm,vei2,lip} which show a double peaked profile on H$\beta$ and [OIII], with an inferred outflow velocity of 
$\sim1000$km s$^{-1}$. As our spectra are of comparable spectral resolution, we would expect to see these wings, if they are present. The 
implications from this are discussed in \S \ref{01003analysis}. The ground-based spectra also show a flat continuum, whereas our optical spectra 
are very blue; this difference can be ascribed to our spectra targeting only the active nuclear regions rather than the whole galaxy. 

Three lines in the I band spectrum, at (observed-frame) 6416\AA, 6535\AA\ and 6941\AA, proved difficult to identify reliably. The line at 6941\AA\ 
has no plausible IDs at 
the systemic redshift (z=0.117). This line could be [FeVII]$\lambda$6087, but this implies a velocity shift of $7000\pm1500$km s$^{-1}$ 
relative to systemic (or a redshift of z=0.141). This shift is, to our knowledge, at least 3000km s$^{-1}$ larger than the shifts seen for this line 
in any other galaxy. Adopting this same shift, then the features at 6416\AA\ and 6535\AA\ would be identified as [FeVI]$\lambda$5632 and 
[FeVII]$\lambda$5721 respectively. In this circumstance the required ratio $[FeVII]\lambda6087/[FeVII]\lambda5721>1.0$ is also satisfied. Equally 
plausible IDs for the features at 6416\AA\ and 6535\AA\ however are CIII$\lambda$5696 and CIV$\lambda$5808 respectively, both signatures of WC type 
Wolf-Rayet stars. In this case the implied velocity shift is a perfectly reasonable $\sim1300$km s$^{-1}$ (or a redshift of z=0.125). Finally, plausible IDs for
the features at 6416\AA\ and 6535\AA\ at the systemic redshift are [NII]$\lambda$5756 and HeI$\lambda$5876. 

These three lines at 6416\AA, 6535\AA\ and 6941\AA\ present a conundrum. All three can be identified as high excitation iron lines at the same redshift, 
but the implied velocity shift is barely believable. Two of the three features can also be identified as lines characteristic of WC type W-R stars, which 
is consistent with the other W-R star signatures seen in the UV and optical spectra. These same two lines can also be identified as commonly seen emission lines 
at the systemic redshift. Given this uncertainty, any conclusions we draw are uncertain. Barring a plausible alternative for the feature at 
6941\AA\ we {\it tentatively} conclude that this feature is [FeVII]$\lambda$6087. This implies that the other two feature may be due at least in part to 
high excitation iron lines, but the presence of NIII$\lambda\lambda\lambda 4634,4640,4642$ also implies that they could be due to W-R stars. Without 
higher spectral resolution data, we can only adopt the most conservative conclusions; the line at 6535\AA\ we conclude is a blend of [FeVII]$\lambda$5721, 
CIV$\lambda$5808 and HeI$\lambda$5876, and the feature at 6416\AA\ we conclude is a blend of [FeVI]$\lambda$5632, CIII$\lambda$5696 and [NII]$\lambda$5756.

A careful search for the more common stellar absorption features (Ca(H)$\lambda$3934 and Ca(K)$\lambda$3968; the `G' band feature at 4250\AA\ 
due to CN and CH; Mg I $b$ $\lambda$5176,5200; Na I$D\lambda5896$, dTiO $\lambda\lambda\lambda$6180,7100,7700; CaII $\lambda$8542) revealed only weak G band 
absorption, and a possible absorption feature from CaII $\lambda$8542. Their weakness or absence in these spectra is not unexpected; our 
observations only include the small part of the host galaxy that lies along the line of sight to the nucleus, and these features are known to be 
diluted by either an AGN power-law continuum, or the continuum from hot young stars.

\subsubsection{Wolf-Rayet stars} \label{01003wr}
The presence of W-R star spectral features allows us to set limits on the stellar populations within the nuclear regions (see e.g. \citet{scv}). 
\citet{ahm} postulate, based on the comparable strength of the NIII $\lambda\lambda\lambda 4634,4640,4642$ lines and the HeII $\lambda 4686$ line, 
and the absence of NV $\lambda 4604$ and  CIV $\lambda\lambda\lambda 4648,5801,5812$, that the W-R stars in F01003-2238 are mainly WN stars, 
with relatively few WC stars. Our spectra are consistent with the presence of a large population of WN stars, but for the reasons given in 
\S\ref{01003prop} we cannot draw reliable conclusions about the presence or otherwise of WC stars solely on the basis of line identifications. Our spectra however 
do not show the oxygen lines that are characteristic of WO stars (OIV $\lambda 3400$, OVI $\lambda\lambda 3811,3834$, \citet{bar}). We therefore 
conclude that at least some of the WR stars in the nucleus of F01003-2238 are of the WN subtype, with some unquantified fraction being WC stars. 
Concerning ourselves solely with the WN stars, then identifying the subtype of WN star is difficult. Again based on the relative strengths of the 
HeII $\lambda 4686$ line and the NIII `bump' at 4600\AA, we postulate that the WN stars in F01003-2238 are probably late type (i.e. WNL stars), in 
agreement with \citet{ahm}, though higher spectral resolution data to resolve the NIII $\lambda\lambda\lambda 4634,4640,4642$ lines would be required 
to confirm or refute this. 

With an identified WR star subtype, the number of WR stars can be estimated, as can the ratio of the number of WR stars to the number 
of O stars (strictly speaking the number of O stars of sufficient mass to evolve into WR stars, typically $\gtrsim30M_{\odot}$). 
We here assume, solely for the purposes of deriving $N_{WR}/N_{O}$, that all the W-R stars are WNL stars.
In the solar neighborhood, $N_{WR}/N_{O}\sim0.1$, a value consistent with continuous star formation. Higher values 
of $N_{WR}/N_{O}$ can only be accommodated if the star formation happens in short duration bursts, or if the metal abundance 
is above Solar. The number of WNL stars can be estimated using the flux of the HeII $\lambda 4686$ line \citep{vac}, assuming that 
all of the HeII $\lambda 4686$ flux is due to WNL stars. Taking the HeII $\lambda 4686$ luminosity of a single WNL 
star to be $1.7\times10^{36}$ergs s$^{-1}$, then for our spectra this gives $1.2\times10^{6}$ or $1.4\times10^{5}$ WNL stars for intrinsic 
Balmer decrements of 2.85 and 3.1 respectively. Computing the ratio of WNL stars to O stars is more complex. The prescription given by 
\citet{vac} for local W-R galaxies assumes pure case B recombination, and a single effective temperature and electron density. The 
maximum value of L$_{obs}(\lambda4686)$/L$_{obs}$(H$\beta)$ from \citet{vac} is 0.21, corresponding to excitation purely by WNL stars, whereas our 
derived value is 0.24. We therefore conclude that our initial assumption is not valid, and that F01003-2238 contains a significant number 
of WC stars in addition to the WN stars. Given the uncertainties in reddening introduced because the H$\alpha$ and [NII]$\lambda\lambda$6548,6583 
lines could not be resolved, and the probable presence of an AGN (\S \ref{01003analysis}), reliable estimates of the ratio of W-R stars to O stars 
are not possible without higher spectral resolution data. Nevertheless, our results imply at least 10$^{5}$ W-R stars in the nucleus of F01003-2238, 
and a value of $N_{WR}/N_{O}$ that is substantially higher than in the Solar neighborhood; this is supportive of a `burst' rather than continuous star formation.

\subsubsection{Analysis} \label{01003analysis}
The I-band spectrum lacks the spectral resolution to reliably deblend the $H\alpha$ and [NII] lines, with variations in the derived $H\alpha$ 
flux of up to a factor of 3 depending on the initial line parameters input to {\it Specfit}. We therefore adopted two canonical values for 
the [NII] $\lambda 6583 / H\alpha$ line ratio, [NII] $\lambda 6583 / H\alpha = 1.0$ (corresponding to an AGN) and [NII] $\lambda 6583 / H\alpha = 0.25$ 
(corresponding to an HII spectrum), and derived line ratio classifications. In both cases, the diagnostics, presented in Table \ref{lineratios} 
and Figure \ref{linediags}, are ambiguous. Other diagnostics do not clarify matters; the 
[OIII] $\lambda 5007 / H\beta$ vs. [OII] $\lambda 3727$/[OIII] $\lambda 5007$ diagnostic classifies the spectrum as an HII region, 
and the [OIII] $\lambda 5007 / H\beta$ vs. [OII] $\lambda\lambda 7320,7330/H\alpha$ diagnostic classifies the spectrum as an AGN, 
irrespective of the assumed intrinsic Balmer decrement. The extinction corrected star formation rate derived using Equation \ref{oiisfr} 
is $6\pm2$M$_{\odot}$ yr$^{-1}$ and $100\pm30$M$_{\odot}$ yr$^{-1}$ for assumed intrinsic Balmer decrements of 2.85 and 3.1 respectively. 

The [SII] $\lambda 6716$ and $\lambda 6731$ lines could not be deblended in this object, hence we only present an estimate of the electron 
temperature in the narrow line region. For an assumed intrinsic Balmer decrement of 2.85 the [OII] $\lambda 3727$ / [OII] $\lambda 7325$ 
ratio is 8.11, whereas for an intrinsic Balmer decrement 3.1 the ratio is 1.7. For a ratio of 1.7, the electron temperatures are $18,700$K, 
$6,600$K, and $5,300$K for assumed electron densities of $n_{e}=2\times10^{4}$cm$^{-3}$, $n_{e}=1\times10^{5}$cm$^{-3}$ and 
$n_{e}=2\times10^{5}$cm$^{-3}$ respectively. For a ratio of 8.11, the electron temperatures are lower, at $12000$K, $8550$K and $6500$K 
for assumed densities of $n_{e}=5\times10^{3}$cm$^{-3}$, $n_{e}=1\times10^{4}$cm$^{-3}$, and $n_{e}=2\times10^{4}$cm$^{-3}$ respectively. 
This indicates that the narrow line region in F01003-2238 does not contain particularly dense gas, and that shock heating does not play 
a significant role. The detection of the [OII] $\lambda 7325$ line is however marginal, at $\sim3\sigma$; these results 
should therefore be regarded with caution. 

The clear signatures of massive star formation in the UV spectrum allow us to examine the starburst activity in F01003-2238 in detail. 
We have modeled the UV spectrum 
using the {\it Starburst99} v4.0 spectral synthesis code \citep{lei1999} using model parameters spanning the following ranges; the mode 
of star formation (i.e. continuous or a `burst'), a starburst age of between 1Myr and 100Myr, an initial mass function slope of between 
2 and 30, and a lower stellar mass limit of between 0.1M$_{\odot}$ and 8M$_{\odot}$. The following analysis assumes that the star formation 
in F01003-2238 can be described by a Salpeter IMF, and that the stellar evolutionary tracks used by {\it Starburst99} are applicable. The 
dereddened UV spectrum, together with the best-fit {\it Starburst99} model, are plotted in Figure \ref{uv_modela}. The W-R star lines, 
together with the strong PCygni profile on the NV line and SiIV doublet, provide strong constraints on the starburst age, limiting the 
age to 3-4Myr (corresponding to the time taken for a 40M$_{\odot}$ O star to evolve into a W-R star). A `burst' of star formation is also 
favored over continuous star formation. The youth of the starburst makes constraining the lower stellar mass limits difficult; we found 
that the data were consistent across the whole 0.1M$_{\odot}$ to 8M$_{\odot}$ range. Constraints can however be set on the slope of the 
IMF: the PCygni profiles, together with the CIII $\lambda$1175 absorption feature, require an IMF slope of 3.3 or less. IMF slopes steeper 
than about 3.3 predict weak or no PCygni profiles or absorption troughs in young starbursts, and can be excluded. There are however two 
significant differences between the observed spectrum and the best-fit model. Firstly, the Ly$\alpha$ emission seen in the spectrum is 
not in the model, as Starburst99 models younger than about 10Myr in general show Ly$\alpha$ in absorption. As the Ly$\alpha$ line is 
broad, with a corrected FWHM of $\sim2100$km s$^{-1}$, we attribute this line to an AGN rather than the starburst (though we note that 
this width is only slightly over the `official', but entirely arbitrary lower limit for line widths in broad line regions of $2000$km s$^{-1}$). A corollary 
to this is that, if high excitation iron lines are present in the I band spectrum, then this would also indicate the presence of an AGN 
(a full discussion of possible origins of the `coronal' iron lines is presented in \S\ref{05189}). Secondly, the SiII$\lambda$1260 and  
CII$\lambda$1335 lines seen in the model are not seen in the spectrum. We attribute this to a combination of lack of instrumental resolution, 
and dilution from the AGN continuum. 

The non-detection of blue wings on the H$\beta$ and [OIII] lines is intriguing, given that such wings are clearly detected in lower 
resolution ground based spectra. The simplest conclusion would be that these blue wings do not exist, however they were detected 
in two independent ground-based measurements \citep{ahm,lip}. This apparent conflict is best explained in terms of different 
apertures; the STIS spectra cover only the central 350pc, whereas the ground-based spectra cover 2kpc or more. If the blue wings 
seen in the ground-based spectra are interpreted as evidence for an outflow, then two possibilities exist. The first is that the 
outflow does not originate in the nucleus. We cannot discount this possibility, however we consider it unlikely. The outflow found 
in the ground-based spectra is almost certainly driven by a starburst or AGN (the other possibility being that it is a dynamical 
relic from the merger); we expect starbursts and AGN to be triggered in the nuclei of ULIRGs, and F01003-2238 is a single nucleus 
system. The second possibility is that this wind was generated some time ago, and that the STIS aperture only covers a small part 
of the wind. Taking a mean velocity of 1000km s$^{-1}$ and assuming a distance of 500pc then this implies that the wind was initiated 
around 500,000 years ago.

\subsection{IRAS F05189-2524} \label{05189}
\subsubsection{Background}
This object is classified as a late-stage merger from WFPC2 and ground-based imaging \citep{sur,vei3}. Ground based 
observations show a Sy2 spectrum \citep{vei}, and what appears to be a single, very red nucleus bisected by 
a dust lane \citep{sur2,sco}. The presence of an obscured AGN was confirmed both by the observations of broad lines in polarized 
flux \citep{you}, and in direct light \citep{vei2}. Later observations \citep{dud} observed the $11.3\mu$m dust feature, 
suggesting that this object also contains an obscured starburst. The 2-10KeV X-ray spectrum \citep{ris} is well fit by 
a two-component model, consisting of a power law with $\Gamma = 1.89^{+0.35}_{-0.34}$, absorbed by a column density of 
$N_{H} = 4.7\times10^{22}$$^{+1.4}_{-1.1}$ cm$^{-2}$, and a thermal component with kT = $0.88^{+0.89}_{-0.35}$ keV. The power 
law component can be interpreted as arising from a Compton thin AGN, with the thermal component either due to the AGN 
or to starburst activity. Results from fitting the $1-1000\mu$m spectral energy distribution of this object \citep{far2} 
found that both a starburst and an AGN were required to explain the infrared emission, with the starburst as the 
major contributor. Conversely, recent X-ray observations \citep{ima} suggest that the starburst is not  
energetically dominant, with the AGN as the probable source of most of the bolometric emission. 

As shown in Figure \ref{slits} the STIS slit was placed across both bright regions in the center of F05189-2524. Consequently, 
two spectra are visible in the STIS image, which are discussed in the following subsections. 

\subsubsection{Knot one}
\begin{figure}
\plotone{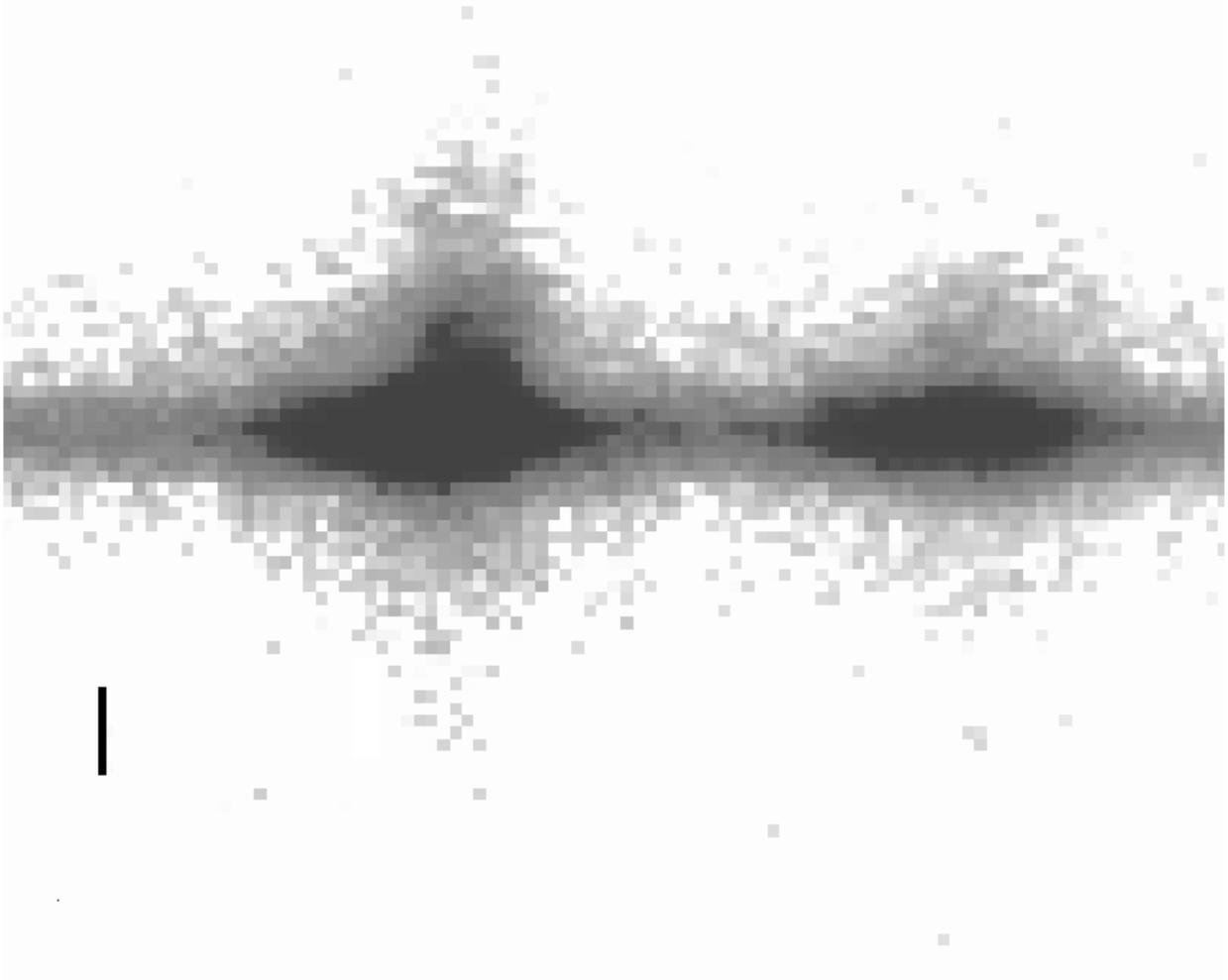}
\caption{Zoom of the STIS MAMA spectrum of F05189-2524, showing the extended, asymmetric Ly$\alpha$ line profile (left) with
the more compact, symmetric NV $\lambda$1240 line. The vertical bar denotes 150pc.  \label{lyaplusnv05189}}
\end{figure}

\begin{figure}
\plotone{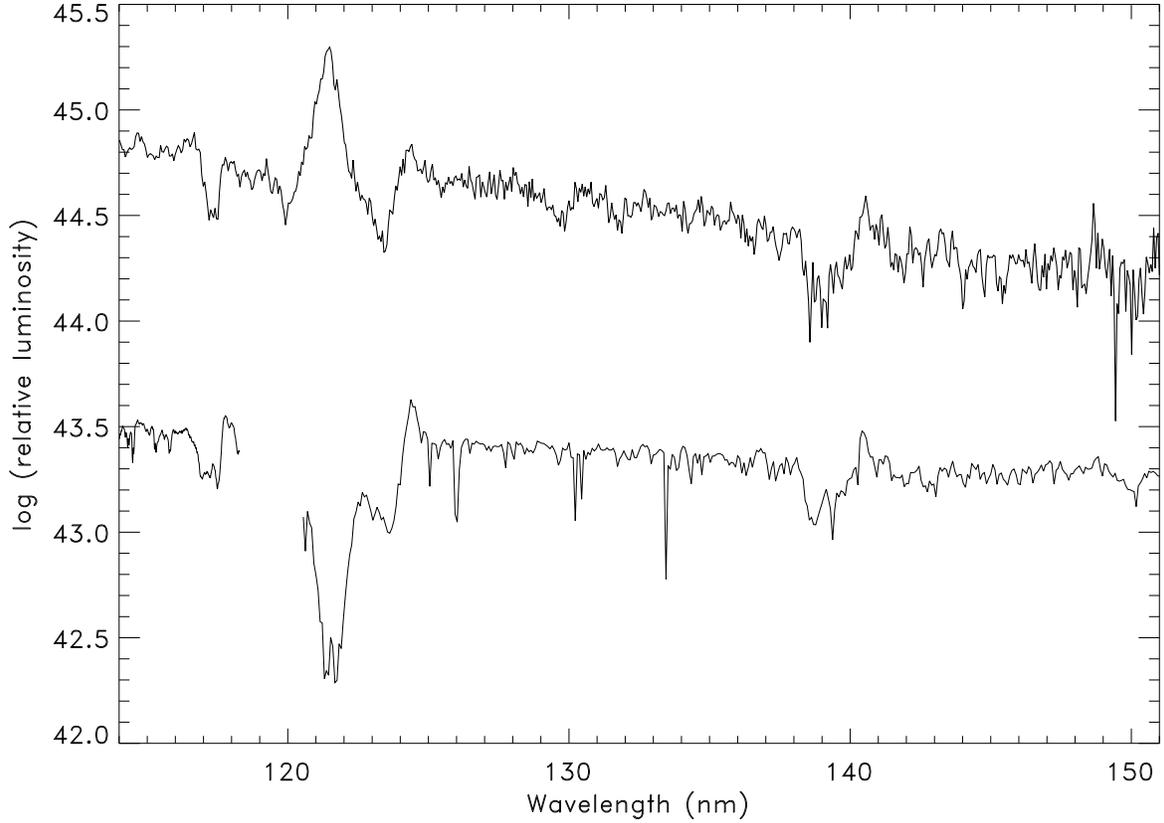}
\caption{Comparison of the UV spectrum of F01003-2238 with a {\it Starburst99} instantaneous burst model with age 3Myr and an IMF slope of 2.0. 
The observed spectrum has been dereddened using Equation \ref{uvextinct} and is plotted in the rest-frame. The model has been offset below the 
spectrum for clarity. The gap in wavelength coverage for the model between 1180\AA and 1200\AA arises because {\it Starburst99} does not produce 
a model spectrum in this wavelength range. \label{uv_modela}}
\end{figure}

\subsubsubsection{Emission line properties}
This knot is the brighter of the two in both the UV and the optical. In the UV several strong emission lines can be seen, 
including Ly$\alpha$ and CIV$\lambda 1549$. Several weaker emission lines are also present, including NIV] $\lambda 1488$. This 
line is not seen in AGN, including composite QSO spectra \citep{vdb} and is thought to originate in winds from Wolf-Rayet stars. 
None of the UV lines show PCygni profiles or evidence for absorption. The Ly$\alpha$ line is spatially extended, and asymmetric 
with respect to the continuum, whereas the spatially compact higher ionization lines, including NV$\lambda$1240, show no asymmetry 
(Figure \ref{lyaplusnv05189}). Asymmetry is also seen in the H$\alpha$, H$\beta$ and [OIII] lines. The spatial extent of the Ly$\alpha$, 
H$\alpha$, H$\beta$ and [OIII] lines is approximately 450 - 500pc, whereas for the higher ionization UV lines the spatial extent 
is 250 - 300pc. The asymmetric region on the Ly$\alpha$ and optical lines is $\sim130$pc across (for F05189-2524 the STIS pixel 
scale is $\sim22$ pc pixel$^{-1}$ in the UV and $\sim50$ pc pixel$^{-1}$ in the optical). We interpret this region as evidence for 
asymmetry in the ionized gas in the nuclear regions, possibly a sign of a late stage merger. In the UV the line widths are broad, 
with a corrected  Ly$\alpha$ FWHM of 2175km s$^{-1}$. In the optical both a narrow and broad component can be seen on the H$\beta$ 
line. None of the optical or UV lines show blue wings, though the broad components of H$\beta$ and [OIII]$\lambda 5007$ show a 
systematic blueshift compared to the other optical lines, corresponding to a velocity of 300-600 km s$^{-1}$ for H$\beta$ and 
900-1200 km s$^{-1}$ for [OIII]$\lambda 5007$. The widths of the `narrow' lines are broad, with corrected FWHMs of order 1000 km s$^{-1}$. 
No W-R star features are visible in the optical spectra. 

In addition to the standard emission lines (Figure \ref{05189_spec} and Table \ref{olinefluxes}), three features are present in 
the B band spectrum between 3050\AA\ and 3350\AA\ that are not commonly seen in starburst or AGN spectra. The features at $\sim3080$\AA\ 
are most likely to be FeII multiplets; as they lie at the extreme edge of the spectrum we do not consider them further. The line at 
3250\AA\ we identify as the OIII $\lambda3133$ resonance-fluorescence line. Our reasoning is  as follows. Adopting a redshift based 
on the Ly$\alpha$ line gives a rest-frame wavelength for this feature of 3131\AA , which is the same within the errors as the wavelength 
of the strongest of the Bowen resonance-fluorescence lines of OIII, at 3133\AA\ . The OIII resonance-fluorescence lines are created in 
nebulae via excitation from HeII Ly$\alpha$ photons with rest-frame wavelength 303.8\AA\ \citep{ost89}, so there is a physical motivation 
for adopting a redshift based on the UV lines rather than the optical lines. The OIII $\lambda$3133 line is also seen in composite QSO 
spectra \citep{vdb}. Finally, we identify the line at 3333\AA\ as being the HeI 3189\AA\ triplet.

\subsubsubsection{High excitation iron lines}
A number of high excitation iron lines, listed in Table \ref{excite}, are present in the optical spectra. These lines show a systematic 
blueshift, corresponding to a velocity of $\sim1200$km s$^{-1}$, and include the very high ionization states [FeX] $\lambda$6375, [FeXI] 
$\lambda$7892 and, remarkably, [FeXIV] $\lambda$5303 (which may be blended with [CaV] $\lambda$5309). The [FeX]-[FeXIV] lines are usually only seen 
in spectra of local supernova remnants, and in the Solar corona (hence their description as `coronal' lines), and are rare in 
extragalactic objects. To our knowledge, only two other galaxy spectra show a direct detection of [FeXIV]; III ZW 77 \citep{ost81} and 
MCG-6-30-15 \citep{rey}. 

Two possible origins have been proposed for these coronal iron lines: (1) they originate in a `Coronal Line Region' (CLR) intermediate 
in distance between the broad and narrow line regions \citep{pen}, (2) they are generated in the ISM several kpc from a `naked' Seyfert 
nucleus \citep{kor}. The second of these possibilities makes predictions regarding relative line luminosities (assuming solar metallicity). 
Specifically, this model predicts that [NeV] $\lambda$3426 will have a comparable line width to [FeX], and be $\sim$12 times stronger, 
and makes further predictions for the ratio of the luminosities of the iron lines to the luminosity of the narrow component of H$\beta$. 
While the widths of [FeX] and [NeV] $\lambda$3426 are comparable (2100$\pm$600km s$^{-1}$ vs. 2500$\pm$800km s$^{-1}$), [NeV] $\lambda$3426 
is approximately 40 times stronger than [FeX] (we corrected the [FeX] 6375 flux from contamination from [OI] 6364 by assuming 
[OI] $\lambda6364$/[OI] $\lambda6300 = \frac{1}{3}$). Furthermore, the L([FeX])/H$\beta$, L([FeXI])/H$\beta$ and L([FeXIV])/H$\beta$ ratios 
predicted by \citet{kor} and those in our spectra all differ by factors of between 2 and 30, depending on the assumed Balmer decrement. 
The first of the models predicts that these iron lines may show a shift (blue or red) compared to the systematic redshift of the system, 
and that the FWHMs of the iron lines will be intermediate between the forbidden lines and the broad components of the permitted lines. 
Both these criteria are met, hence we conclude that the high excitation iron lines in F05189-2524 originate in a CLR between 
the narrow and broad line regions.  

Under the assumption that these high excitation iron lines are produced in photoionized gas (rather than shocks in radio jets), then the lines 
can be used to constrain the temperature and density of this gas within a few pc of the central ionizing source. The detection of [FeXIV] implies \citep{fkf} 
a range in Hydrogen column density of $5.75<$log n(H)(cm$^{-2})<8.0$, and a peak electron temperature of $\sim160,000$K. The detection of 
this line also implies that we have a line of sight to regions very close to the central ionizing source; if the hydrogen column density 
range and electron temperature are correct then [FeXIV] is expected to be produced only within $\sim0.2$pc of the broad line region.

\subsubsubsection{Analysis}
The complex lines of sight to the AGN intimated by the presence of the high ionization iron lines makes deriving reliable optical line 
diagnostics complex. The $H\beta$ line cannot be reliably fitted with more than a single Gaussian and single Lorentzian component. Attempts 
to fit the $H\alpha$ line with more than a single component were similarly unsuccessful; when a broad and narrow component were simultaneously 
fitted then the flux of the broad component varied by $\sim30\%$, whereas the flux of the narrow component varied by up to an order of magnitude, 
depending on the values of the input parameters to {\it specfit}. We therefore assumed, solely for the purposes of deriving line 
diagnostics, [NII] $\lambda 6583 / H\alpha$ (narrow) = 1.0. For the $H\beta$ line we used only the Gaussian component when measuring 
the [OIII] $\lambda 5007 / H\beta$. The  $H\beta$ flux we used is thus likely too low as it does not include any contribution from a narrow 
Lorentzian component. The value of  log([OIII] $\lambda 5007 / H\beta$) plotted in Figure 7 is thus probably too high, and indeed a value of 
1.56 is higher than any other published AGN.  This overestimate however should not affect the classification. The three primary line ratio 
diagnostics all classify this object as an AGN, as do the [OIII] $\lambda 5007 / H\beta$ vs. [OII] $\lambda 3727$/[OIII] $\lambda 5007$ and 
[OIII] $\lambda 5007 / H\beta$ vs. [OII] $\lambda\lambda 7320,7330/H\alpha$ diagnostics. Furthermore, as the I band spectrum shows the 
[SIII] $\lambda\lambda 9069,9531$ lines we were able to employ the [SIII] $\lambda\lambda 9069,9531 / H\alpha$ vs. [OII] $\lambda\lambda 7320,7330/H\alpha$ 
diagnostic described in \citet{otv}, which also classifies this spectrum as an AGN. 

With an assumed intrinsic Balmer decrement of 3.10, the extinction corrected star formation rate derived using Equation \ref{oiisfr} is 
$30\pm10$M$_{\odot}$ yr$^{-1}$; this should of course be regarded as an upper limit due to the presence of an AGN. The [SII] $\lambda 6716$ 
\& [SII] $\lambda 6731$ lines could not be reliably deblended in this object, hence we only present an estimate of the electron temperature 
in the narrow line region. For an assumed intrinsic Balmer decrement of 3.1 the [OII] $\lambda 3727$ / [OII] $\lambda 7325$ ratio is 16.98, 
which predicts an electron temperature of $13,900$K for an assumed electron density of $n_{e}=1\times10^{3}$cm$^{-3}$, and $6,200$K for an 
assumed electron density of $n_{e}=1\times10^{4}$cm$^{-3}$. 

The strong, broad emission lines in the UV spectrum evidently arise from an AGN, however the analysis we could perform was 
limited, for two reasons. Firstly, most emission line diagnostics for the ionizing source in the narrow line region of AGN use one or 
both of the OVI $\lambda1034$ and CIII] $\lambda$1909 lines \citep{ham99}, and both these lines lie outside the G140L bandpass at $z=0.043$. Secondly, due 
to the limited spectral resolution of our data it proved impossible to separate the broad and narrow components of most of the 
emission lines in the UV; the only line this was achieved for was CIV $\lambda$1549 and even in this case the 1$\sigma$ errors on the resulting 
widths and fluxes were $\sim30\%$. We therefore first used the [OIII]$\lambda 5007$/H$\beta$ vs. 
[NeV]$\lambda$3426/[NeIII]$\lambda$3869 diagnostic \citep{adt} to distinguish between shocks and photoionization, which implied a 
photoionizing continuum with $\alpha\sim -1$. We then used the NIV]/CIV diagnostic presented by \citet{ham02} and the {\it total} 
CIV$\lambda$1549 flux, to derive a lower limit to the metallicity, which gave log$(Z/Z_{\odot})>0.4$. 

Overall, F05189-2524 knot 1 clearly contains an AGN, based on both the line shapes in the optical and UV and the optical line diagnostics. 
Furthermore, we have a line of sight very close to the central regions of this AGN, based on the broad component on H$\beta$ and 
the detection of the high excitation `coronal' iron lines. Although there are no unambiguous signs of a starburst in the optical spectra, 
the presence of NIV] $\lambda$1488 in the UV suggests a large population of W-R stars, implying that a young (3-4Myr) starburst is 
also present.

Finally, we note that an interesting experiment would be to measure the polarization of the high-ionization lines found in 
this spectrum. The broad components found on the H$\alpha$ and H$\beta$ lines in our spectra, and results from previous authors 
\citep{you,vei2} imply that the BLR in this knot is moderately, though not completely, obscured. As the distance of the source of the 
high-excitation iron lines from the BLR is predicted to decrease with increasing ionization stage \citep{fkf}, the relative polarizations 
of these lines would be an interesting diagnostic of the change in obscuration as a function of distance from the BLR.

\subsubsection{Knot two} \label{05189_knot2}
This knot is fainter than knot one, and is not detected in the UV. Due to the spatially extended emission lines in knot 1, which 
contaminate most of the key diagnostic lines in knot 2, little quantitative information can be derived. Nevertheless, a simple 
comparison of the optical spectra of knots 1 and 2 shows two clear differences. Knot 2 shows a distinct Balmer break and Na ID 
$\lambda$5896 absorption, neither of which is seen in knot 1. Furthermore, knot 2 does not show the HeII $\lambda$4686 line; based 
on the H$\beta$ and [NeIII] $\lambda$3869 fluxes of the two knots we would have expected to detect HeII $\lambda$4686 in knot 2, if it 
were present. We speculate that knot 2 has a higher obscuration level than knot 1, and somewhat lower ionization, though our data 
do not allow us to quantify this. Overall, the properties of this knot combined with the properties of knot 1 are not consistent 
with the central region of F05189-2524 being a single nucleus bisected by a dust lane, contrary to the results of \citet{sur2} and 
\citet{sco}. Instead, the most likely interpretation is that F05189-2524 is a single nucleus system, and that knot 2 is a second 
line emitting region ionized by the nucleus in knot 1, but with a different level of excitation and obscuration. 

One puzzling point is the clear detection of a line we identify as [FeX] 6375 in knot 2. Careful inspection of the 2-d image shows that 
this line does not arise via contamination from knot 1, but if it does arise from knot 2 then the apparent lack of other high excitation 
lines is strange. The simplest scenario which fits with a single nucleus interpretation is that the [FeX] emission from knot 2 is a result 
of scattering and reflection from knot 1, in which case we would expect this line to be polarized, but we cannot quantify this or explain 
the absence of the other high excitation iron lines without higher quality data.

\subsection{IRAS F12071-0444}\label{12071}
HST imaging shows a central, reddened, compact nucleus surrounded by a `chaotic' high surface brightness region and a number of 
small blue knots, and two tidal tails \citep{sur}. Ground based optical spectroscopy shows a Sy2 spectrum \citep{vei,kew}, and 
\citet{vei97} use near-infrared spectroscopy to infer the presence of a hidden BLR.

A single spectrum is evident in the STIS B and I band frames, with nothing detected in the UV frame. The optical spectra contain 
several high excitation Neon lines and the MgII $\lambda\lambda$2796,2803 doublet. Both H$\gamma$ and H$\delta$ are seen in emission. 
The optical lines are symmetric about the continuum, with spatial extents of 800 - 1000pc (for F12071-0444 the STIS pixel scale is $\sim120$ pc 
pixel$^{-1}$ in the optical). Two high excitation iron lines are also present in the I band spectrum, $[$Fe VII$]$ $\lambda$5721 
and $\lambda$6087 (Table \ref{excite}), detected at the $\sim4\sigma$ level. All the line ratio diagnostics classify this 
spectrum as an AGN, and as a Sy2 rather than a LINER. The [SII] $\lambda\lambda 6716,6731$ doublet was not resolved for this object, 
hence we only estimated electron temperatures in the narrow line region. The [OII] $\lambda 3727$ / [OII] $\lambda 7325$ ratio, at 55.58, 
predicts a temperature in the NLR of $9700$K for $n_{e}=100$cm$^{-3}$, and $7200$K for $n_{e}=1000$cm$^{-3}$. Overall 
we classify this object as an AGN, with no clear signs of an accompanying starburst, though we cannot rule out the presence of a starburst 
in the other knots in this object. 

\subsection{IRAS F15206+3342}\label{15206}
\citet{san3} classify this galaxy as a Sy2 on the basis of ground based optical spectroscopy. \citet{vei97} do not however 
find any evidence for a hidden BLR. HST and ground-based imaging shows a complex morphology, with a large number of knots and 
a tidal tail \citep{sur,vei3}. Most of these knots are probably starbursts, but two may be active nuclei. \citet{aco} find evidence 
for gas inflow along the tidal tail, and postulate that this object is a late stage merger between two disk galaxies which suffered 
a retrograde encounter. 

For this object the STIS slit was oriented so that it covered the centers of two knots, one of which was identified as a possible AGN by 
\citet{sur}, and partially covered a third (Figure \ref{slits}). Three spectra are thus visible in the STIS frames, and are discussed 
in the following subsections.

\subsubsection{Knot one}

\subsubsubsection{Emission \& absorption lines}
Knot 1 is the brightest of the three in the UV, and the UV spectrum contains a number of absorption systems, including Ly$\alpha$. 
Only two emission lines are present; NV $\lambda$1240 and CIII $\lambda$1175. The (unsaturated) absorption systems have FWHMs of 
800 - 1500 km s$^{-1}$ and show a blueshift of $\sim300$km s$^{-1}$. In addition to the commonly seen UV absorption lines there are 
two weak features at observed-frame 1598\AA\ and 1609\AA\ that proved difficult to identify; features at these wavelengths are (to 
our knowledge) not seen in spectra of AGN or starburst galaxies, or composite QSO spectra, and do not correspond to absorption features 
commonly seen in Damped Lyman Alpha systems \citep{pro01}. These two features are most likely spurious, and so are not marked in Figure 
\ref{15206_587_spec}, even though they are detected at $\sim3.5\sigma$. If they are real, then the most likely candidates, based on 
currently available atomic line list data, are multiplets of allowed and semi-forbidden transitions of Fe II and Fe III (another 
possible ID motivated by absorption features seen in DLAs for the longer wavelength feature is PbII $\lambda$1433), but it is beyond 
the scope of this paper to provide a secure identification.

The optical spectra show a steep continuum and a Balmer break. The emission lines, which are offset from the continuum by $\sim300$pc, 
have a spatial extent of 800 - 900pc. We detect both the [NeV] $\lambda$3426 and [NeIII] $\lambda$3869 lines, but no high excitation 
iron lines. A number of absorption features are present blueward of 4000\AA; we identify MgII $\lambda\lambda$2796,2803 and (cautiously) 
MgI $\lambda$2852, HeI $\lambda$2945, FeII $\lambda$3180 and FeI+FeII $\lambda$3261. These absorption features show blueshifts of between 
$300$km s$^{-1}$ (for MgII $\lambda\lambda$2796,2803) and $2600$km s$^{-1}$ (for FeI+FeII $\lambda$3261) relative to the systemic redshift 
as measured from [OII]$\lambda 3727$. The optical emission line widths are narrow, with a corrected [OIII]$\lambda 5007$ FWHM of 
$620\pm50$ km s$^{-1}$. 

\subsubsubsection{Analysis}
It proved impossible to disentangle the H$\alpha$ and [NII] emission; attempts to do so produced errors of at least a factor of two on 
the derived H$\alpha$ flux depending on the input paramaters to {\it specfit}, hence the same procedure as for IRAS F01003-2238 was used 
to derive line ratio diagnostics. There is some ambiguity in the classification of this spectrum. The [OIII] $\lambda 5007 / H\beta$ vs. 
[NII] $\lambda 6583 / H\alpha$ and [OIII] $\lambda 5007 / H\beta$ vs. [SII] $\lambda\lambda 6716,6731/H\alpha$ diagnostics classify 
this knot as either an AGN or a starburst, depending on the assumed intrinsic Balmer decrement. The weak upper limit on 
[OI] $\lambda 6300$ renders diagnostics based on this line inconclusive. The [OIII] $\lambda 5007 / H\beta$ vs. [OII] $\lambda\lambda 7320,7330/H\alpha$ 
and [OIII] $\lambda 5007 / H\beta$ vs. [OII] $\lambda 3727$/[OIII] $\lambda 5007$ diagnostics classify this object as an AGN and HII 
region respectively, irrespective of any assumptions. We do not present estimates for electron densities or temperatures in the NLR 
as the [OII] $\lambda 7325$ line is not detected, and the [SII] doublet could not be deblended. With an assumed intrinsic 
Balmer decrement of 2.85, the extinction corrected star formation rate (if all the [OII] $\lambda 3727$ flux arises from star formation) 
derived using Equation \ref{oiisfr} is $60\pm15$M$_{\odot}$ yr$^{-1}$. 

\begin{figure}
\plotone{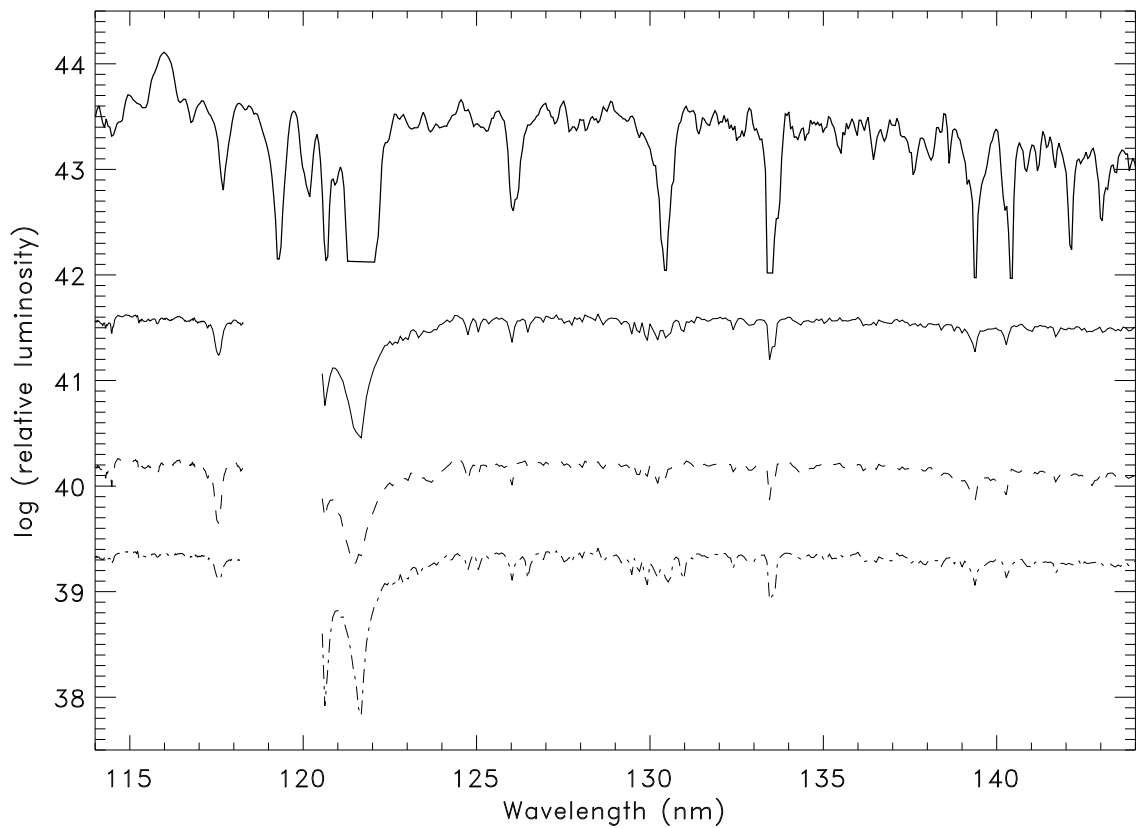}
\caption{Comparison of the UV spectrum of F15206+3342 knot 1 (top) with representative {\it Starburst99} spectral synthesis models. 
A 6Myr burst with an IMF slope of 2.35 (dashed line) is a good fit to the NV emission and a reasonable fit to the Ly$\alpha$ absorption profile, 
but cannot explain the other absorption features. A 20Myr burst with an IMF slope of 3.3 (dot-dashed line) is a good fit to the Ly$\alpha$ 
absorption, but cannot explain the  NV emission and is a very poor match to the absorption features. Even combining the two models (solid line) 
does not produce a good fit. Note we are {\it not} invoking multiple starbursts in this spectrum, the models are shown simply for comparison. 
\label{uv_modelb}}
\end{figure}

To quantify the presence of a starburst and/or AGN within this knot, we have modeled the UV spectrum using the {\it Starburst99} v4.0 
code (\citet{lei1999} and the same parameter ranges as for F01003-2238 (\S \ref{01003analysis}), under the {\it ab initio} assumption 
that the UV spectrum is starburst dominated. The observed UV spectrum is not consistent with a single episode or any combination of two 
episodes of star formation (Figure \ref{uv_modelb}), irrespective of the adopted age, IMF slope and lower stellar mass limit, or whether the star formation is 
continuous or a `burst'. The emission line we identify as CIII 1175\AA\ is not present (this line is in fact not predicted to exist in 
{\it any} of the starburst models we investigated), and the multiple strong absorption features in the observed spectrum are poorly 
matched by the model. Significant absorption at 1465\AA\ due to OI $\lambda 1304$ and SiII $\lambda 1307$ could be 
generated by a starburst with a very steep IMF slope ($\gtrsim7$) but then the other absorption lines, particularly Ly$\alpha$, would 
not be consistent with the data. A close match around the Ly$\alpha$ and NV lines can however be obtained with a single starburst model. 
The remainder of the UV spectrum requires a different explanation.

There are two possible origins for the absorption features in the UV spectrum of F15206+3342. The first is that they 
arise due to interstellar absorption. Interstellar absorption features in the rest-frame UV that are qualitatively similar to those 
seen in our spectrum are seen in, for example, UV spectra of Lyman Break Galaxies \citep{pet00,sha03}. It is possible therefore that the 
features in our spectrum are due to interstellar absorption. An argument against this however 
is that most of the absorption features in our spectrum are between 2 and 5 times broader than those seen in Lyman Break Galaxy spectra. 
For example, for the SiIV $\lambda\lambda 1394,1403$ doublet; Pettini et al find rest-frame widths of 1.94\AA\ and 1.42\AA, Shapley et 
al find widths of 1.83\AA\ and 0.81\AA, whereas we find widths of 4.65\AA\ and 3.71\AA. The exception is the CII $\lambda 1335$ 
line, which has a comparable width in our spectrum and that of Shapley et al. 

The second possible origin for these features is that they arise all or in part from a Broad Absorption Line (BAL) region, itself an 
outflow driven by a hidden AGN. Approximately 10\% of QSOs show BALs with widths $\gtrsim2000$km s$^{-1}$ 
\citep{wey91,ara01,hal02,lac02,rei03}. The existence of an AGN driven BAL outflow in F15206+3342 is supported by the presence of MgII 
$\lambda\lambda$2796,2803 and other absorption lines seen in BALs \citep{hal02} in the B band spectrum, which would classify this object as a low 
ionization BAL, or `LoBAL', which comprise $\sim10$\% of the BAL population. The UV and B band 
spectra contain absorption features that are thought to arise from metastable excited levels of iron; such features are very rare 
amongst LoBALs, and would place F15206+3342 knot one in the so-called `FeLoBAL' class of AGN \citep{haz87,bec97,lac02}. This picture 
however is also problematic, for two reasons. Firstly, the absorption lines are narrow compared to the classical definition 
of absorption lines in a BAL QSO \citep{wey91}; the corrected FWHMs of MgII $\lambda\lambda$2796,2803 and SiIV $\lambda$1394 in our spectra are 
$1500\pm150$km s$^{-1}$ and $1100\pm200$km s$^{-1}$, though \citet{hal02} do find absorption troughs of comparable widths to ours in some 
SDSS BAL QSOs. Our object would thus be formally termed a `mini-FeLoBAL' AGN \citep{ham01}. Secondly, a BAL nucleus cannot be responsible 
for all the observed features in the UV spectrum. BAL QSOs show absorption in CIV, NV, SiIV and Ly$\alpha$ in order of {\it decreasing} 
strength (\citet{hal02} \& refs therein) whereas our spectrum shows strong Ly$\alpha$ absorption and weak NV emission. Overall therefore, 
we conclude that the most likely scenario is that the Ly$\alpha$ absorption and NV emission arise from a starburst, and the UV and optical 
absorption features arise from an outflow driven by a mini-FeloBAL AGN, but we cannot discount the possibility that some or all of the 
absorption features (particularly CII $\lambda1335$) arise due to interstellar absorption of a starburst wind. 

If the BAL AGN interpretation is correct then it would have interesting implications for the origin of the BAL QSO phenomenon. There are 
currently two models for the origin of the absorption lines in BAL QSOs; the `disk wind' model and the `youth' model. In the disk wind model 
\citep{muc98,psk00} a wind from an accretion disk encounters a high column density, highly ionized gas. The high ionization state of this 
gas means it can transmit UV photons from (and thus be driven outwards by) resonance lines such as CIV $\lambda 1549$, but the high column 
density of the gas shields it from higher energy photons that would otherwise completely ionize the gas. In this model 
BAL QSOs are those QSOs viewed along a line of sight that coincides with this outflowing gas, marking the detection of BALs 
in QSOs as largely orientation dependent \citep{elv00}. The `youth' model on the other hand contends that BAL QSOs are young 
objects, still surrounded by large quantities of gas and dust; in this case the BALs do not arise due to a particular line of sight 
\citep{vwk93,bec97,wbp99}. The `youth' model is often applied to LoBAL QSOs; \citet{spr92} propose that the differences between the line 
properties of LoBAL QSOs and those of non-BAL QSOs cannot be explained solely in terms of different relative orientations.
\citet{vwk93} propose an alternative scenario in which LoBALs form via ablation of dust by UV photons in outflows arranged either as a thick disk 
or as an isotropic distribution of clouds; in this case LoBALs do fit within AGN orientation schemes, and this scenario is consistent with polarimetric 
observations of large samples of BAL QSOs \citep{sch,ogl,huts}. Even for this model, \citet{vwk93} note that BAL QSOs would likely evolve into 
less obscured QSOs as the outflowing material either settles into a thinner configuration or entirely ablates away, and propose that this scenario 
is consistent with LoBAL QSOs being youthful objects. An evolutionary link between LoBAL QSOs and ULIRGs has also been previously suggested 
from other observations \citep{lip94,can01}. The presence of a BAL nucleus within F15206+3342 would add further credence to the `youth' 
model for LoBAL QSOs.

\subsubsection{Knot two}
Knot 2 is the faintest of the three in the optical, and the UV emission is also very weak. This is due at least in part to the 
positioning of the slit; as can be seen from Figure \ref{slits} the slit passes over only one edge of this knot. The discussion 
of this spectrum should therefore be regarded with caution. In the optical, similar emission lines are present as in knot 1, 
including the [NeIII] $\lambda$3869 line, but also the H$\gamma$ and HeI $\lambda$5876 lines. The [NeV] $\lambda$3426 line may 
be double peaked; adopting the systemic redshift then the rest-frame wavelengths of both features, at 3396\AA\ and 3422\AA\ lie 
just within the range seen in (for example) SDSS QSOs \citep{vdb}, but we conservatively only identify the longer wavelength 
feature as [NeV] $\lambda$3426. No stellar absorption features were reliably identified.

It proved impossible to disentangle the $H\alpha$ and [NII] emission, hence the same procedure was used as for knot 1. The 
results from our three primary diagnostics are ambiguous between an HII and AGN, as is the [OIII] $\lambda 5007 / H\beta$ vs. 
[OII] $\lambda 3727$/[OIII] $\lambda 5007$ diagnostic, though the [OIII] $\lambda 5007 / H\beta$ vs. [OII] $\lambda 3727$/[OIII] 
$\lambda 5007$ diagnostic robustly classifies this object as an HII region. Overall therefore, we classify this knot as a starburst.

The [OII] $\lambda 7325$ line is not detected in this object, however the [SII]$\lambda\lambda 6716,6725$ doublet was successfully 
deblended, hence we present estimates for the electron density in the NLR. The [SII] $\lambda 6716$ / [SII] $\lambda 6731$ ratio, at 
0.90, infers electron densities of 1200cm$^{-3}$, 910cm$^{-3}$, and 690cm$^{-3}$ for temperatures of 5,000K, 10,000K and 20,000K
respectively, which implies a moderately dense NLR. With an assumed intrinsic Balmer decrement of 2.85, the optical star formation 
rate derived using Equation \ref{oiisfr} is $75\pm20$M$_{\odot}$ yr$^{-1}$ 

\subsubsection{Knot three}\label{15206k3}
Knot 3 is optically the brightest of the three, but is not detected at all in the UV. The lines are symmetric, with a spatial extent 
of $\sim850$pc. The line widths are narrow compared to the other spectra in the sample, with a corrected [OIII] $\lambda 5007$ FWHM of 
$450\pm50$ km s$^{-1}$. Though the continuum detection was strong in both the B and I band spectra, no stellar absorption features were 
identified.

For this object the lines are sufficiently narrow that the $H\alpha$ and [NII] lines can be reliably deblended. All five of our 
adopted line diagnostics classify this spectrum as a starburst rather than an AGN. Both the [OII] $\lambda\lambda 3727,7325$ 
doublets are detected, and the [SII]$\lambda\lambda 6716,6725$ doublet was successfully deblended. From the 
[OII] $\lambda 3727$ / [OII] $\lambda 7325$ ratio, 31.53, we derive electron temperatures of 13900K, 9300K and 5100K for densities of 
10$^{2}$cm$^{-3}$, 10$^{3}$cm$^{-3}$ and 10$^{4}$cm$^{-3}$ respectively. From the [SII] $\lambda 6716$ / [SII] $\lambda 6731$ ratio 
of 0.77 we derive electron densities of 1200cm$^{-3}$, 1600cm$^{-3}$ and 2100cm$^{-3}$ for assumed temperatures of 5000K, 10000K and 
20000K respectively. 

As this knot is universally classified as an HII region, and because the [NII] and H$\alpha$ lines could be deblended, we 
have estimated the oxygen abundance using the [NII] $\lambda$6584 /[OII] $\lambda\lambda$3726,3729 diagnostic presented by \citet{kdo}, 
though we note that this diagnostic is only reliable for metallicities above one half solar. We derive:

\begin{equation}
log \left(\frac{[NII] \lambda 6584}{[OII] \lambda\lambda 3726,3729}\right) = -0.35, 
\end{equation}

\noindent which corresponds to an oxygen abundance of:

\begin{equation}
 log(O/H) + 12 = 8.91 
\end{equation}

This oxygen abundance is higher than solar (8.69, \citet{aba}) and is comparable to the most metal rich local UV selected
galaxies (e.g. \citet{cont}). 

With an assumed intrinsic Balmer decrement of 2.85, the extinction corrected star formation rate derived using Equation \ref{oiisfr} 
is $1000\pm300$M$_{\odot}$ yr$^{-1}$. Even by the standards of local ULIRGs, this star formation rate is extraordinarily high \citep{far2}, 
and is comparable to that seen in distant HLIRGs \citep{far4}. If we instead use the extinction corrected H$\alpha$ luminosity, and 
the calibration given by \citet{ken}:

\begin{equation}
SFR = 7.9\times10^{-42}L(H\alpha) \label{hasfr}
\end{equation}

\noindent then we obtain a comparable value; $800\pm250$M$_{\odot}$ yr$^{-1}$. Even if all the star formation in F15206+3342 was 
in this knot and visible in the optical, this star formation rate is likely to be too high by a few hundred M$_{\odot}$ yr$^{-1}$.
Establishing whether or not this is the case, and the reasons behind any overestimate, is beyond the scope of this paper, but we 
briefly mention two possibilities. As noted by \citet{ken}, star formation rate calibrations for H$\alpha$ and [OII]$\lambda 3727$ 
depend strongly on both the optical extinction, and the assumed stellar IMF. It is quite possible that the measured extinction in 
this knot is incorrect. It is also possible that the IMF slope is steeper than that assumed in deriving Equations \ref{oiisfr} and \ref{hasfr}, 
this would mean there are a larger number of high-mass stars present, producing a greater ionising flux per M$_{\odot}$. 

The unambiguous classification of this knot as an HII region means that an AGN is unlikely to contribute significantly to the continuum 
emission. We can therefore estimate the stellar mass of this knot by combining the extinction measure from our observations with 
diagnostics from HST WFPC2 imaging \citep{sur}; we note this is the only knot in our sample for which this analysis is possible, as the 
others are either weakly detected or contain an AGN. This gives a mass of $\sim2\times10^{8}$M$_{\odot}$ (note this differs from the mass 
presented in Table \ref{knots} as we have applied a reddening correction based on our spectra). This translates, assuming the knot 
is approximately spherical, to a stellar mass density of $\sim200$M$_{\odot}$pc$^{-3}$. Both the mass and stellar density of this knot 
are comparable to, but at the lower ends of, the ranges seen in the cores of elliptical galaxies, which typically have masses up to 
$\sim10^{10}$M$_{\odot}$ and stellar densities in the range $10^{2} - 10^{3}$M$_{\odot}$ \citep{lau89}. Coupled with the very high star 
formation rate (at least a few hundred M$_{\odot}$ yr$^{-1}$) this makes this knot an excellent candidate for the forming core of an 
elliptical galaxy.

\section{Summary \& Conclusions}\label{conc}
We have presented HST STIS longslit optical and UV spectra of the knots in the nuclear regions of four `warm' ULIRGs, and used a 
combination of line ratio diagnostics and spectral synthesis modelling to constrain the active power source in each knot. A summary 
of the properties of the knots in each object is as follows:

\noindent 1) {\it IRAS F01003-2238} - The STIS data show a single bright narrow line spectrum. The UV spectrum contains signatures of 
O supergiants, both WC and WN type Wolf-Rayet stars, and (probably) an AGN. Results from spectral synthesis modelling are consistent with 
a `burst' of star formation with age 3-4Myr and an IMF slope $\lesssim3.3$. The optical spectra show what may be high excitation iron 
lines, indicating a line of sight close to the broad line region, and evidence for a wind, probably starburst driven, that originated 
in the nucleus around 0.5Myr ago. 

\noindent 2) {\it IRAS F05189-2524} - Two spectra can be seen, corresponding to the two sections of the central bisected `nucleus'. The UV 
spectrum of knot 1 shows several strong permitted lines from an AGN, together with weak NIV] $\lambda$1488 emission that probably arises 
from a young ($\sim4$Myr) `burst' of star formation. The B and I band spectra of knot 1 show a large number of lines, including the OIII 
$\lambda$3133 Bowen resonance-fluorescence line. Also present are several high excitation iron lines, including [FeXIV] $\lambda$5303, 
which imply lines of sight from the narrow line region through to $\sim0.2$pc from the accretion disk. The optical 
line diagnostics classify this knot as an AGN. Further diagnostics show no evidence for shock heating, and imply a metallicity 
of 2.5Z$_{\odot}$ or higher. Knot 2 is fainter than knot 1, and is not detected in the UV. Its properties are consistent with a higher 
obscuration and lower ionization level than knot 1. Considered together, the properties of the two knots are not consistent with a single 
nucleus bisected by a dust lane, but rather with a single nucleus (knot 1) together with a second knot of line emitting gas ionized by the 
nucleus in knot 1 but with different ionisation conditions and obscuration.

\noindent 3) {\it IRAS F12071-0444} - A single spectrum of the nucleus can be seen. This object is undetected in the UV, and the continuum is 
only weakly detected in the B and I bands. Only a few optical emission lines are present, limiting the scope of the analysis. The optical 
line ratio diagnostics universally classify this object as a Sy2. There is no evidence for an accompanying starburst, though we cannot rule 
out the possiblility of starbursts in the other knots in this object. 

\noindent 4) {\it IRAS F15206+3342} - Three spectra can be seen. The UV spectrum of knot 1 is a chimera; containing Ly$\alpha$ and NV 
lines that can be modeled as arising from a starburst, together with a CIII $\lambda$1175 line that is not present in any spectral synthesis 
model we investigated, and thus likely arises from an AGN. The UV and B band spectra contain a number of absorption systems that 
probably arise from a hidden FeLoBAL AGN, the alternative being that they arise from interstellar absorption. Knot 2 is only weakly detected; 
this spectrum is classified as an HII region with a moderately dense narrow line region. Knot 3 is optically bright but not detected in the UV. Line 
ratio diagnostics classify this knot as an HII region with a supersolar (Z$\sim$1.5Z$_{\odot}$) metallicity and a very high star formation 
rate. Combined with previous results this implies that this knot is plausibly the forming core of an elliptical galaxy. This is the only 
object in the sample for which the knot classifications given by \citet{sur} were found to be incorrect, for knots 1 and 3. 

The knots in our sample of 4 `warm' ULIRGs are remarkable both for their similarities and their differences. Three of the four objects contain 
a knot with clear AGN signatures, and one object (F01003-2238) also likely contains an AGN, demonstrating that selection based on 'warm' mid-IR 
colors does preferentially find ULIRGs with an AGN. In three out of four cases this AGN was found in the knot identified as a putative AGN from 
multiband HST imaging. Three of the four ULIRGs in our sample also contain at least one starburst knot, confirming that most ULIRGs are `composite' 
objects, containing both a starburst and an AGN. The detection of luminous AGN and/or starbursts in 
all of the knots in our sample constitutes plausible, though not compelling evidence that these knots are the sites for the dust-shrouded starburst 
and AGN activity that power the IR emission in ULIRGs. The AGN display a wide range of properties; two are classical narrow line AGN, one shows 
both broad and narrow lines and has multiple lines of sight to the center, and one is a narrow line AGN with (plausibly) an FeLoBAL outflow. Of 
the starbursts, only 3 have constraints on their ages, with two having ages 3-4Myr and the third with an age of $\lesssim20$Myr. Though this 
could be a selection effect, it seems that young bursts of star formation are common in ULIRGs, indicating that most ULIRGs probably undergo 
several such bursts during their lifetime. Finally, the properties of one starburst knot are consistent with it being the forming core of an 
elliptical galaxy, the first time to our knowledge that such a structure has been identified in a ULIRG. The identification of such a structure 
is supportive of the idea that ULIRGs can evolve into systems similar to local elliptical galaxies. 

Our study has demonstrated the power of combining high spatial resolution optical and UV spectroscopy to study local ULIRGs; the optical data, 
in addition to providing lines for standard line ratio diagnostics, also isolated several rare, high excitation lines in most systems. In several 
cases the AGN was only unambiguously seen in the UV, and quantitative constraints on the properties of the starbursts only came from modelling 
the UV spectra. In one case (F01003-2238) this approach showed (probable) AGN signatures, when such signatures had previously only been seen in 
mid-IR spectroscopy. While this method does not directly measure what contributes to the IR emission it does give a sensitive, largely model free 
measure of what {\it could} contribute to the IR emission. Indeed, this study would not have been possible without a space-based optical/UV 
spectrograph with spatial resolution of $\lesssim 0.1 \arcsec$. Future multiwavelength studies focused on these knots are likely to give powerful 
insights into the nature of starburst and AGN activity in ULIRGs, and into ULIRG evolution. 

\acknowledgments
We are grateful to J. Afonso, L. Hartley, M. Lacy, D. C. Kim and H. E. Smith for helpful discussion, and to the referee for a very helpful report. 
The data presented here were obtained using the NASA/ESA {\em Hubble Space Telescope}, obtained at the Space Telescope Science Institute, which 
is operated by the Association of Universities for Research in Astronomy, Inc., under NASA contract NAS 5-2655. Support for proposal number 
GO-08190 was provided by NASA through a grant from the Space Telescope Science Institute, which is operated by the Association of Universities 
for Research in Astronomy, Incorporated, under NASA contract NAS5-26555. The research described in this paper was carried out, in part, by the 
Jet Propulsion Laboratory, California Institute of Technology, and was sponsored by the National Aeronautics and Space Administration. SV 
acknowledges partial support of this research from NASA under grant NASA/LTSA NAG-56547. DBS acknowledges support from a Senior Award from the 
Alexander von Humboldt Foundation and from the Max Planck-Institut fur extraterrestrische Physik. This research has made use of the NASA/IPAC 
Extragalactic Database (NED) which is operated by the Jet Propulsion Laboratory, California Institute of Technology, under contract with the 
National Aeronautics and Space Administration.

\begin{deluxetable}{lccccr} 
\tablecolumns{6} 
\tablewidth{0pc} 
\tablecaption{`Warm' ultraluminous infrared galaxy sample \label{sample}} 
\tablehead{ 
\colhead{Galaxy}&\colhead{RA (J2000)}&\colhead{Dec}&\colhead{Redshift}&\colhead{$L_{ir}$\tablenotemark{a}}&\colhead{$m_{B}$}
}
\startdata 
IRAS F01003-2238 & 01 02 49.99 & -22 21 57.5  & 0.1177 & 12.39 & 18.5 \\ 
IRAS F05189-2524 & 05 21 01.47 & -25 21 45.4  & 0.0426 & 12.20 & 15.6 \\ 
IRAS F12071-0444 & 12 09 45.12 & -05 01 13.9  & 0.1284 & 12.51 & 18.2 \\ 
IRAS F15206+3342 & 15 22 38.05 & +33 31 35.9  & 0.1244 & 12.33 & 16.8 \\ 
\enddata
\medskip

$^{a}$Logarithm of the $8-1000\mu$m luminosity, taken from \citet{ks98} and rescaled to our cosmology.
\end{deluxetable}

\begin{deluxetable}{lccccc} 
\tablecolumns{6} 
\tablewidth{0pc} 
\tablecaption{Knot magnitudes, radii and masses \label{knots}} 
\tablehead{ 
\colhead{Galaxy}&\colhead{Knot}&\colhead{$m_{B}$}&\colhead{$m_{I}$}&\colhead{Radius} &\colhead{Mass} \\
\colhead{      }&\colhead{    }&\colhead{       }&\colhead{       }&\colhead{pc    } &\colhead{M$_{\odot}$\tablenotemark{b}}
}
\startdata 
     F01003-2238 & 1\tablenotemark{a} & 19.27 & 18.32 & 28  & $3\times10^{8}$    \\
     F05189-2524 & 1\tablenotemark{a} & 18.78 & 16.55 & 25  & $2\times10^{10}$   \\ 
                 & 2                  & 19.01 & 17.37 & 27  & $1\times10^{9}$    \\ 
     F12071-0444 & 1\tablenotemark{a} & 21.59 & 19.00 & 244 & $4\times10^{10}$   \\ 
     F15206+3342 & 1                  & 19.45 & 18.54 & 88  & $3\times10^{9}$    \\ 
                 & 2                  & 19.50 & 18.68 & 104 & $2\times10^{9}$    \\ 
                 & 3\tablenotemark{a} & 19.29 & 17.37 & 57  & $3\times10^{10}$\tablenotemark{c}  \\ 
\enddata 
\tablecomments{All data are taken from \citet{sur}. Knots are numbered as in Figure \ref{slits}.}
\tablenotetext{a}{knots identified by \citet{sur} as putative AGN based on their broad-band colors.}
\tablenotetext{b}{stellar mass inferred by \citet{sur} assuming that all of the B and I band flux is due to stellar emission}
\tablenotetext{c}{but see \S\ref{15206k3}}
\end{deluxetable}

\begin{deluxetable}{lccccc} 
\tablecolumns{6} 
\tablewidth{0pc} 
\tablecaption{Ultraviolet emission line fluxes \label{ulinefluxes}} 
\tablehead{ 
\colhead{Line}&\colhead{Wavelength}&\colhead{F01003-2238}&\colhead{F05189-2524}&\colhead{F12071-0444}&\multicolumn{1}{c}{F15206+3342} \\
\colhead{    }&\colhead{\AA    }&\colhead{  1   }&\colhead{  1  }&\colhead{  1   }&\colhead{1}
}
\startdata 
HeII                        & 1085      & 1.19          & --                    &  --           & --           \\ 
CIII \tablenotemark{a}      & 1175      &  abs          & 0.13:                 &  --           & 1.23:        \\ 
Ly$\alpha$                  & 1216      & 10.83         & 52.74                 &  --           & abs          \\ 
NV                          & 1240      & 1.58:         & 20.51                 &  --           & 0.22:        \\ 
SiII                        & 1260      &  abs          & --                    &  --           & abs          \\ 
CII                         & 1335      &  --           & 0.90                  &  --           & abs          \\ 
SiIV                        & 1394,1403 & 1.08:         & 2.74                  &  --           & abs          \\ 
NIV$]$                      & 1488      & 0.53:         & 1.12                  &  --           & --           \\ 
CIV                         & 1548,1551 &   --          & 28.54                 &  --           & --           \\ 
$[$NeV$]$ \tablenotemark{a} & 1575      &   --          & 0.30::                &  --           & --           \\ 
$[$NeIV$]$\tablenotemark{a} & 1601      &   --          & 0.35::                &  --           & --           \\ 
HeII                        & 1640      &   --          & 13.22                 &  --           & --           \\ 
\enddata
\tablecomments{Fluxes are in units of $10^{-15}$ergs cm$^{-2}$s$^{-1}$ and have not been corrected for 
extinction. Flux errors are of the order $5\% - 10\%$ for most sources; those fluxes marked with a colon have 
flux errors of $\sim20\%$ and those fluxes with a double colon have errors of $\sim30\%$.  } \tablenotetext{a}{Uncertain ID}
\end{deluxetable} 

\begin{deluxetable}{lcccccccc} 
\tablecolumns{9} 
\tablewidth{0pc} 
\tablecaption{Optical emission line fluxes \label{olinefluxes}} 
\tablehead{ 
\colhead{Line}  & \colhead{Wavelength}&\colhead{F01003-2238}&\multicolumn{2}{c}{F05189-2524}&\colhead{F12071-0444}&\multicolumn{3}{c}{F15206+3342} \\
\colhead{    }  & \colhead{\AA    }   &\colhead{ 1    }&\colhead{1}&\colhead{2\tablenotemark{a}}&\colhead{  1   }&\colhead{1}&\colhead{2}&\colhead{3}
}
\startdata 
MgII                & 2796,2803      &  5.13         & N/A     & N/A       & 0.42:         & abs       & --        &  --       \\ 
MgI                 & 2852           &  --           & N/A     & N/A       & --            & abs       & --        &  --       \\ 
FeII                & 2985           &  --           & N/A     & N/A       & --            & abs       & --        &  --       \\ 
HeI                 & 2945           &  --           & N/A     & N/A       & --            & abs       & --        &  --       \\ 
OIII                & 3133           &  --           & 7.13    & N/A       & --            & --        & --        &  --       \\ 
FeII                & 3180           &  --           & --      & N/A       & --            & abs       & --        &  --       \\ 
HeI                 & 3189           &  --           & 5.26:   & N/A       & --            & --        & --        &  --       \\ 
FeI+FeII            & 3261           &  --           & --      & N/A       & --            & abs       & --        &  --       \\ 
$[$NeV$]$           & 3346           &  --           & 21.96   & N/A       &  --           & --        & --        &  --       \\ 
$[$NeV$]$           & 3426           &  2.68:        & 58.44   & 0.72:     & 0.39:         & 0.41:     & 0.34:     & 0.55      \\ 
$[$OII$]$           & 3727           &  1.27:        & 3.91:   & --        & 0.71          & 1.70      & 1.55      & 2.01      \\ 
$[$NeIII$]$         & 3869           &  6.09         & 17.11   & 0.38:     & 1.16          & 0.52::    & 0.16::    & 1.14      \\ 
$[$NeIII$]$         & 3968           &  2.26         & 5.50    & abs       & 0.37:         & --        & --        & --        \\ 
$[$SII$]$           & 4068,4076      &   --          & 1.75:   & --        & --            & --        & --        & --        \\ 
H$\delta$           & 4102           &  3.18::       & 2.09    & abs       & 0.41::        & --        & --        & 0.63:     \\ 
H$\gamma$           & 4340           &  2.38         & 5.51    & --        & 0.40:         & --        & 0.43      & 1.84      \\ 
NIII                & 4634,4640,4642 &  1.26:        &  --     & --        & --            & --        & --        & --        \\ 
HeII                & 4686           &  1.23::       & 15.92   & --        & 0.64          & --        & --        & --        \\ 
H$\beta$            & 4861           &  4.96         & 34.12   & 0.23:     & 1.42          & 0.95      & 1.25      & 5.49      \\ 
$[$OIII$]$          & 4959           &  9.07         & 62.44   & N/A       & 6.91          & 1.28      & 1.32      & 8.37      \\ 
$[$OIII$]$          & 5007           &  23.67        & 177.98  & N/A       & 20.82         & 4.20      & 4.31      & 21.43     \\ 
HeII                & 5412           &   --          & 5.72    & --        &  --           & --        & --        & --        \\ 
HeI                 & 5876           &   --\tablenotemark{b}& 4.28    & --        &  --           & --        & 0.35:     & 2.66:     \\  
$[$OI$]$            & 6300           &  3.23         & 4.37    & --        & 1.97          & --        & --        & --        \\ 
$[$NII$]$           & 6548           &   --          & 13.30   & N/A       & 6.17          & --        & --        & 6.54      \\ 
H$\alpha$           & 6563           &   --          & 202.20  & N/A       & 13.99         & --        & --        & 68.26     \\ 
$[$NII$]$           & 6583           &   --          & 39.94   & N/A       & 18.54         & --        & --        & 19.66     \\ 
H$\alpha$+$[$NII$]$ &  --            & 56.94         & --      & N/A       &  --           & 8.76      & 12.49     & --        \\
$[$SII$]$           & 6716,6731      & 3.29          & 12.95   & 0.51:     & 5.94          & 1.71:     & 1.15      & 3.76      \\ 
HeI                 & 7065           &   --          & 1.47:   & --        &  --           & --        & --        & --        \\
$[$ArIII$]$         & 7138           &   --          & 5.14    & --        & 1.29          & --        & --        & 1.96::    \\ 
$[$OII$]$           & 7320,7330      & 2.28:         & 2.86:   & --        & 0.58:         & --        & --        & 2.27:     \\ 
$[$SIII$]$          & 9069           &   --          & 17.22   & N/A       &  --           & --        & --        & 8.35::    \\ 
$[$SIII$]$          & 9531           &   --          & 25.08:  & N/A       &  --           & --        & --        & --        \\ 
\enddata
\tablecomments{Fluxes are in units of $10^{-15}$ergs cm$^{-2}$s$^{-1}$ and have not been corrected for extinction. 
`--': Undetected. `N/A': line is either outside the bandpass, contaminated by companion 
spectra, or in a noisy part of the spectrum. Flux errors are of the order $5\% - 10\%$ for most sources; those fluxes marked with a colon have 
flux errors of $\sim20\%$ and those fluxes with a double colon have errors of $\sim30\%$.} 
\tablenotetext{a}{extracted using a 1 pixel ($42$pc) aperture, fluxes are uncertain due to contamination from knot 1.}
\tablenotetext{b}{but see notes for Table \ref{excite}}
\end{deluxetable}

\begin{deluxetable}{lcccccc} 
\tablecolumns{7} 
\tablewidth{0pc} 
\tablecaption{Corrected Full Widths at Half Maximum of selected emission lines \label{linewidths}} 
\tablehead{ 
\colhead{Line}      &\colhead{F01003-2238} &\colhead{F05189-2524}   &\colhead{F12071-0444} &\multicolumn{3}{c}{F15206+3342} \\
\colhead{    }      &\colhead{ 1    } &\colhead{ 1   }   &\colhead{  1   } &\colhead{1}  &\colhead{2} &\colhead{3}
}
\startdata 
Ly$\alpha$          & $2200\pm130$                 & N/A              &  --            & abs         & --         & --          \\ 
CIV 1549            &     --                       & $4600\pm1000$(B) &  --            & --          & --         & --          \\ 
                    &     --                       & $1000\pm500$ (N) &  --            & --          & --         & --          \\ 
$[$OII$]$ 3727      & $860\pm500$                  & $900\pm275$      & $500\pm100$    & $600\pm300$ & $600\pm50$ & $550\pm100$ \\ 
H$\beta$            & $2100\pm170$                 & $3700\pm900$ (B) & $525\pm100$    & $600\pm100$ & $500\pm50$ & $450\pm50$  \\ 
                    &     --                       & $800\pm100$  (N) &  --            & --          & --         & --          \\ 
$[$OIII$]$ 5007     & $2400\pm70$                  & $2000\pm300$ (B) & $950\pm50$     & $650\pm50$  & $550\pm50$ & $450\pm50$  \\ 
                    &     --                       & $900\pm50$   (N) &  --            & --          & --         & --          \\ 
$[$FeXIV$]$ 5303    &     --                       & $4800\pm500$     &  --            & --          & --         & --          \\ 
$[$Fe VII$]$ 5721   & $900\pm500$\tablenotemark{a} & $3000\pm200$     & $900\pm700$    & --          & --         & --          \\ 
$[$Fe VII$]$ 6087   & $900\pm500$\tablenotemark{a} & $2200\pm100$     & $400\pm700$    & --          & --         & --          \\ 
$[$OI$]$ 6300       & $2200\pm400$                 & $950\pm300$      & $1100\pm150$   & --          & --         & $600\pm300$ \\ 
$[$Fe X$]$ 6375     &     --                       & $2000\pm600$     & $800\pm350$    & --          & --         & --          \\ 
\enddata 
\tablecomments{Widths are given in km s$^{-1}$. FWHM of Ly$\alpha$ and NV cannot be measured in F05189-2524 as the broad and 
narrow components of these two lines cannot be reliably deblended, see text for details.} 
\tablenotetext{a}{but see notes for Table \ref{excite}}
\end{deluxetable}

\begin{deluxetable}{lcccccccccccc} 
\tablecolumns{13} 
\tablewidth{0pc} 
\tablecaption{Optical line ratios and spectral classifications \label{lineratios}} 
\tablehead{ 
\colhead{Galaxy} & \colhead{} & \colhead{$\frac{H\alpha}{H\beta}$} & \colhead{$A_{V}$} & \colhead{$\frac{[OIII]}{H\beta}$} & \colhead{$\frac{[OII]3726}{[OIII]}$} & \colhead{$\frac{[NII]}{H\alpha}$} & \colhead{$\frac{[SII]}{H\alpha}$} & \colhead{$\frac{[OI]}{H\alpha}$} & \colhead{$\frac{[OII]7325}{H\alpha}$}
 & \colhead{(1)} & \colhead{(2)} & \colhead{(3)}
}
\startdata 
     F01003-2238 & 1 &  8.61 & 3.46 & 4.77  & 0.05: & {\bf 0.25} & 0.08  & 0.05    & 0.05:      & --  & --  & --  \\
                 &   &  2.85 & 0.0  & 4.15  & 0.20: & {\bf 0.25} & 0.07  & 0.06    & 0.04:      & HII & HII & AGN \\
     F01003-2238 & 1 &  4.92 & 1.45 & 4.77  & 0.05: & {\bf 1.0}  & 0.13  & 0.09    & 0.09:      & --  & --  & --  \\
                 &   &  3.10 & 0.0  & 4.50  & 0.09: & {\bf 1.0}  & 0.13  & 0.10    & 0.08:      & AGN & HII & AGN \\
     F05189-2524 & 1 & 35.08 & 7.87 & 39.08 & 0.02: & {\bf 0.25} & 0.08  & 0.03    & 0.02:      & --  & --  & --  \\ 
                 &   &  2.85 & 0.0  & 28.42 & 0.46: & {\bf 0.25} & 0.07  & 0.04    & 0.01:      & AGN & AGN & AGN \\
     F05189-2524 & 1 &  8.77 & 3.26 & 39.08 & 0.02: & {\bf 1.0}  & 0.32  & 0.11    & 0.07:      & --  & --  & --  \\ 
                 &   &  3.10 & 0.0  & 34.25 & 0.08: & {\bf 1.0}  & 0.30  & 0.12    & 0.05:      & AGN & AGN & AGN \\
     F12071-0444 & 1 &  9.87 & 3.63 & 14.69 & 0.09  & 1.33       & 0.42  & 0.14    & 0.04       & --  & --  & --  \\ 
                 &   &  3.10 & 0.0  & 12.68 & 0.38  & 1.31       & 0.39  & 0.16    & 0.03       & AGN & AGN & AGN \\
     F15206+3342 & 1 &  6.88 & 2.76 & 4.40  & 0.40  & {\bf 0.25} & 0.26  & $<$0.18 & $<$0.11    & --  & --  & --  \\ 
                 &   &  2.85 & 0.0  & 3.93  & 1.17  & {\bf 0.25} & 0.25  & $<$0.09 & $<$0.08    & HII & A   & A   \\
                 & 1 &  3.93 & 0.75 & 4.40  & 0.40  & {\bf 1.0}  & 0.46  & $<$0.20 & $<$0.20    & --  & --  & --  \\ 
                 &   &  3.10 & 0.0  & 4.27  & 0.54  & {\bf 1.0}  & 0.45  & $<$0.20 & $<$0.18    & AGN & AGN & A   \\
                 & 2 &  7.50 & 3.03 & 3.45  & 0.36  & {\bf 0.25} & 0.12  & $<$0.03 & $<$0.03    & --  & --  & --  \\ 
                 &   &  2.85 & 0.0  & 3.05  & 1.16  & {\bf 0.25} & 0.11  & $<$0.04 & $<$0.02    & HII & HII & HII \\
                 & 2 &  4.29 & 1.02 & 3.45  & 0.36  & {\bf 1.0}  & 0.21  & $<$0.06 & $<$0.06    & --  & --  & --  \\ 
                 &   &  3.10 & 0.0  & 3.31  & 0.53  & {\bf 1.0}  & 0.21  & $<$0.06 & $<$0.05    & AGN & HII & A   \\
                 & 3 &  12.44& 4.62 & 4.40  & 0.08  & 0.29       & 0.06  & $<$0.05 & 0.03       & --  & --  & --  \\ 
                 &   &  2.85 & 0.0  & 3.65  & 0.49  & 0.28       & 0.05  & $<$0.02 & 0.02       & HII & HII & HII \\
\enddata 
\tablecomments{(1) [OIII] $\lambda 5007 / H\beta$ vs. [NII] $\lambda 6583 / H\alpha$ classification.
(2) [OIII] $\lambda 5007 / H\beta$ vs. [SII] $\lambda\lambda 6716,6731/H\alpha$ classification.
(3) [OIII] $\lambda 5007 / H\beta$ vs. [OI] $\lambda 6300 / H\alpha$ classification.
`HII': HII/Starburst spectrum. `AGN': AGN spectrum. `A': ambiguous classification between an HII and AGN spectrum. 
For F01003-2238 and F15206+3342 knots 1 \& 2 the [NII] and H$\alpha$ lines could not be deblended, and for F05189-2524 the 
narrow and broad components on H$\alpha$ could not be deblended, hence two sets of line ratio diagnostics have been derived 
for these three objects assuming intrinsic Balmer decrements of 2.85 and 3.10 respectively. In these cases the $\frac{[NII]}{H\alpha}$ 
ratio is shown in bold. Uncertainties on the line ratios are $\sim10\%$, except for those ratios marked with 
a colon, where the uncertainty is $\sim20\%$.}
\end{deluxetable}

\begin{deluxetable}{lccccc} 
\tablecolumns{6} 
\tablewidth{0pc} 
\tablecaption{High excitation iron lines \label{excite}} 
\tablehead
{ 
\colhead{Line}   &\colhead{wavelength}&\colhead{F01003-2238}&\multicolumn{2}{c}{F05189-2524}&\colhead{F12071-0444} \\
\colhead{     }  &\colhead{\AA       }&\colhead{  1   }&\colhead{1}&\colhead{2\tablenotemark{a}}&\colhead{1}
}
\startdata 
$[$Fe VI$]$                 & 5530               & --                     & 2.07:     &   --              &  --             \\ 
$[$Fe VI$]$                 & 5632               & 1.95\tablenotemark{c}  & 3.66      &   --              &  --             \\ 
$[$Fe VII$]$                & 3582               & --                     & 3.32      &   --              &  --             \\ 
$[$Fe VII$]$                & 3760               & --                     & 6.39:     &   --              &  --             \\ 
$[$Fe VII$]$                & 5159               & --                     & 16.15     &   --              &  --             \\ 
$[$Fe VII$]$                & 5721               & 1.5:\tablenotemark{d}  & 19.96     &   --              & 0.46:           \\ 
$[$Fe VII$]$                & 6087               & 1.00:\tablenotemark{e} & 31.76     &   --              & 0.58:           \\ 
$[$Fe X$]$\tablenotemark{b} & 6375               & --                     & 9.44      &  0.53             &  --             \\ 
$[$Fe XI$]$                 & 7892               & --                     & 7.85      &   --              &  --             \\ 
$[$Fe XIV$]$                & 5303               & --                     & 18.58     &   --              &  --             \\ 
\enddata
\tablecomments{Line fluxes are in units of $10^{-15}$ergs cm$^{-2}$s$^{-1}$ and have not been corrected for extinction. Flux errors 
are of the order $5\% - 10\%$ for most sources; those fluxes marked with a colon have flux errors of $\sim20\%$.}
\tablenotetext{a}{extracted using a 1 pixel ($\sim42$pc) aperture, IDs and fluxes are uncertain due to contamination from knot 1, see 
\S\ref{05189_knot2} for more details.}
\tablenotetext{b}{corrected for contamination from [OI] $\lambda6364$ assuming [OI] $\lambda6364$/[OI] $\lambda6300 = \frac{1}{3}$.}
\tablenotetext{c}{Uncertain ID, may be blended with CIII$\lambda$5696, [NII]$\lambda$5756}
\tablenotetext{d}{Uncertain ID, may be blended with CIV$\lambda$5808, HeI$\lambda$5876}
\tablenotetext{e}{Uncertain ID}
\end{deluxetable}

\end{document}